\documentclass{report}

\usepackage{url}
\usepackage{graphicx}

\usepackage{amsfonts}
\usepackage{multicol}
\usepackage{flushend} 
\usepackage{caption}
\usepackage{subcaption}
\usepackage{booktabs}
\usepackage{textcomp}
\usepackage{comment}

\usepackage{threeparttable}
\usepackage{titlesec}
\usepackage{setspace}
\usepackage[shortlabels]{enumitem}
\usepackage{stfloats}
\usepackage[below]{placeins}
\usepackage[utf8]{inputenc}
\usepackage{cite}

\usepackage{pifont}

\usepackage{amssymb}

\usepackage{multirow}
\usepackage{amsmath}
\usepackage{csquotes}
\usepackage{color,soul}

\usepackage{comment}
\date{}

\begin{document}

\author{{Ijaz Ahmad}}

\title{PhD Thesis \\IMPROVING SOFTWARE DEFINED COGNITIVE AND SECURE NETWORKING}

\newpage

\maketitle

\noindent UNIVERSITY OF OULU GRADUATE SCHOOL; \\
UNIVERSITY OF OULU, FACULTY OF INFORMATION TECHNOLOGY AND ELECTRICAL ENGINEERING; \\
CENTRE FOR WIRELESS COMMUNICATIONS; \\
INFOTECH OULU \\

\vspace{10mm}

\noindent Supervised by:\\
Prof. Mika Ylianttila\\
Prof. Andrei Gurtov\\

\vspace{10mm}

\noindent Reviewed by:\\
Professor Ivan Ganchev\\
Professor Niklas Carlsson\\

\vspace{10mm}

\noindent Opponent:\\
Professor Pekka Toivanen
\vspace{10mm}

\noindent ISBN 978-952-62-1950-9 (Paperback)\\
ISBN 978-952-62-1951-6 (PDF)\\

\noindent ISSN 0355-3213 (Printed)\\
ISSN 1796-2226 (Online)\\

\vspace{10mm}

\noindent Date of PhD thesis defence: 20.06.2018

\newpage 

\doublespacing

\begin{abstract}

Traditional communication networks consist of large sets of vendor-specific manually configurable devices. These devices are hardwired with specific control logic or algorithms used for different network functions. The resulting networks comprise distributed control plane architectures that are complex in nature, difficult to integrate and operate, and are least efficient in terms of resource usage. However, the rapid increase in data traffic requires the integrated use of diverse access technologies and autonomic network operations with increased resource efficiency. Therefore, the concepts of Software Defined Networking (SDN) are proposed that decouple the network control plane from the data-forwarding plane and logically centralize the control plane. The SDN control plane can integrate a diverse set of devices, and tune them at run-time through vendor-agnostic programmable Application Programming Interfaces (APIs). 

This thesis proposes software defined cognitive networking to enable intelligent use of network resources. Different radio access technologies, including cognitive radios, are integrated through a common control platform to increase the overall network performance. The architectural framework of software defined cognitive networking is presented alongside the experimental performance evaluation. Since SDN enables applications to change the network behavior and centralizes the network control plane to oversee the whole network, it is highly important to investigate SDN in terms of security. Therefore, this thesis finds the potential security vulnerabilities in SDN, studies the proposed security platforms and architectures for those vulnerabilities, and presents future directions for unresolved security vulnerabilities. Furthermore, this thesis also investigates the potential security challenges and their solutions for the enabling technologies of 5G, such as SDN, cloud technologies, and virtual network functions, and provides key insights into increasing the security of 5G networks.

\end{abstract}



\chapter{Introduction}

\section{Background}

Communication networks are extending the notion of connectivity by linking diverse sets of devices and services to each other via the Internet. New things enabled by the Internet of Things (IoT) such as smart home appliances, wearable devices, sensors, and autonomous cars, as well as online services in various fields, including medical ICT, agriculture, transportation, logistics, and defense will utilize communication networks to further improve the quality of human life. Future networks, therefore, need to be equipped with the ability to handle numerous, even though challenging, requirements. For instance, IoT provides the foundational infrastructure for future services, but the foundation of IoT lies in smart networks~\cite{AHMAD201882}. Smart networks use software and network infrastructure abstractions to configure heterogeneous network environments automatically in order to fulfill the requirements of future services~\cite{vandenDam2013}. 

However, traditional communication networks have many challenges in terms of complexity, adaptability, and the user or node mobility~\cite{7509389}. In traditional networks, policies that dictate the network operations are mostly implemented with low-level device configurations. Most of the devices in communication networks use complex control protocols that have large sets of tunable parameters. Vendor-specific manual configurations of a huge number of networked devices, and tuning the control parameters in large networks, excessively complicate the operations of networks~\cite{6385040}~\cite{6461195}. Furthermore, these configurations are mostly prone to human-errors~\cite{Kim:2011:ENC:2068816.2068863}, leading to 50 to 80 percent of the network downtime, the reason being the systems' complexities~\cite{Juniper}. 

Carrier networks usually rely on vendor-specific hardware platforms. Hence, network operators end up in vendor lock-ins due to vendor-specific network control platforms and lack of interoperability between equipment from different vendors~\cite{6553677}. Since the traffic demands of users and requirements of services are always changing, the inevitable mix and match of network equipment from different vendors is becoming a major challenge for network operators. Furthermore, control platforms embedded into network hardware are difficult to upgrade and result in distributed network control architectures. Such distributed control architectures lack global resource visibility and increase network complexity. It is demonstrated in~\cite{benson2009unraveling} that complex network architectures are more prone to outages due to the increased complexity in network configurations.

The amalgamation of an increasing number of services and devices envisioned by, for example, IoT will not only further complicate network management but will also require openness to innovations in order to incorporate new services and solutions. Next generation networks, such as 5G (5th Generation), aim to provide very high data rates with extremely low latency, significantly improved Quality of Service (QoS), and accommodate the proliferation of new and emerging services, as well as an enormous number of devices in mainstream networks~\cite{7414384}. Moreover, 5G networks will overcome the limitations of previous generations of wireless networks in terms of integrating low power and low data rate devices~\cite{8141874}. Therefore, the 5G architecture will not be an incremental advancement of 4G, but a paradigm shift towards very high carrier frequencies with massive bandwidths, extreme Base Stations (BSs) and device densities, and will utilize a diverse set of access technologies~\cite{6824752}. 

5G, therefore, must use different access technologies in a well-coordinated fashion and without cell-centric designs~\cite{6736745}. The Next Generation Mobile Networks (NGMN) alliance~\cite{alliance20155g} also considers cooperative and user-centric design of multiple Radio Access Technologies (RATs) as one of the key requirements of 5G. However, such use of diverse access technologies will be highly challenging in traditional distributed network control architectures. In order to cooperate, the distributed control architectures will synchronize their functions through extra control signaling, whereas the signaling traffic is increasing 50 percent faster than the data traffic~\cite{6957145}. For a network a connecting massive number of IoT devices, the signaling traffic originated by distributed control architectures will be a huge challenge~\cite{6736745}.

Software Defined Networking (SDN) logically centralizes the network control architecture, relinquishes the need of device-level configurations~\cite{Kirkpatrick:2013:SN:2500468.2500473}, breaks the barrier of vendor lock~\cite{7035546}, and opens communication networks for innovation~\cite{mckeown2008openflow}. SDN separates the network control from the data forwarding elements. Thus, the data forwarding elements are rendered in a more simplified form that can be (re)-programmed through a vendor-agnostic interface such as the OpenFlow protocol. By implementing the control platforms in software modules and introducing programmable Application Programming Interfaces (APIs) in network equipment, SDN facilitates fast network features development and deployment~\cite{6461195}. With the logically centralized control platforms overlooking the data forwarding elements, SDN facilitate run-time network manipulation through programmable APIs and ensures coherent network-wide policy implementation. Using the centralized control platform, SDN relinquishes the need for manual per-device configurations and thus reduces the network complexity~\cite{Kirkpatrick:2013:SN:2500468.2500473}. 

The concepts of cognitive networking~\cite{1542652} have been proposed to enable autonomous network operation. Cognitive networks have the capabilities to observe a user's needs, sense the operating environment, and adjust itself accordingly to fulfill the user's needs in that environment. A Cognitive Radio Network (CRN), implementing the radio part of cognitive networking, has the capability to sense free or occupied radio resources (e.g. frequency spectra) and thus enable intelligent cooperative use of the resources. A full cognitive network also needs the upper layers to be adjustable at run-time, and thus, Software Adaptable Network (SAN) elements are required to tune the network at run-time through software. For example, in spectrum management and mobility functions, all the layers, including application, transport, network, medium access control and physical layers are required to cooperate~\cite{AKYILDIZ20062127}. 

However, traditional networks have some challenges that remained as the main barriers in achieving full cognition in an entire network. First, the vertical integration of networked functions that requires all layers to cooperate~\cite{1542652} is one example of those challenges. The distributed control architectures due to the layering intricacies have been further complicating cognition throughout a network~\cite{AKYILDIZ20062127}. Second, the vendor-specific manually configurable network devices that require human intervention in the wake of required changes, in other words, unavailability of SAN elements before the inception of SDN or its most valuable implementation, i.e. OpenFlow is another example of such challenges. Since, SDN resolves the challenges associated with traditional layered architectures and device configurations, integrating the concepts of SDN and cognitive networking opens new frontiers for robust and autonomic network operation and management.

%
%
%


%


\section{Objectives and scope of the thesis} 

The ultimate goal of this thesis is to investigate the potential of SDN in future networks in terms of integrating different access technologies, intelligently sharing resources among the access technologies, and analysis of security in SDNs and SDN-based future networks.
 
Decoupling the network architecture from the infrastructure facilitates innovations in each, since it breaks the dependency of one on the other, and thus enables cost-effective network operation, and feature updates and deployment~\cite{Raghavan:2012:SIA:2390231.2390239}. SDN abstracts network functions from the infrastructure by decoupling the network control and data forwarding planes. By logically centralizing the network control platform and introducing programmability in networks, SDN also facilitates autonomic network management~\cite{8063402}, and highly flexible network operation~\cite{6461195}. Therefore, SDN has the potential to provide the robust communication architecture for future networks and online services~\cite{8017556},~\cite{6957145}.

In traditional networks, each BS or eNodeB (eNB) performs independent resource allocation, scheduling and base band processing, which is not feasible for the vision of future cellular networks~\cite{Gudipati:2013:SSD:2491185.2491207}. The backhaul on the other hand comprises devices that run on vendor-specific manual configurations, have embedded control logic using algorithms to route, control and monitor traffic. The tight coupling of the control and data forwarding planes makes these devices extremely rigid to changes and updates, and the distributed control architecture of wireless networks has resulted in lack of global visibility of network resources making it difficult to deploy network-wide coherent policies. However, next generation wireless networks such as 5G must use a combination of RATs to meet the growing demands of user traffic and traffic generated by IoTs~\cite{6812298}. Therefore, the first objective of this thesis is investigating the potential of SDN to enable centralized control of heterogeneous RATs to meet the future traffic requirements of diverse devices and services.

The growth in wireless systems' capacity can be attributed to the increased use of radio spectrum, which will further increase in future wireless networks~\cite{6736747}. However, a major challenge for future wireless networks will be to efficiently meet the increasing demands for higher network capacity while the spectrum resource remains scarce~\cite{7321983}. One of the key opportunities in fully utilizing the available heterogeneity is efficiently sharing spectrum among RATs~\cite{6568922}. By centralizing the network control, SDN offers the opportunity to fully utilize diversity in RATs by effectively sharing spectrum among them at run-time. Hence, the second objective of this thesis is to investigate dynamic spectrum sharing through the centralized control framework provided by SDN. The concepts of cognitive networking, that use cognitive radios for intelligently sharing spectrum, have been combined with SDN to enable cognition from the physical layer up-to the application layer.

SDN will be the key technology in the next generation networks (e.g. 5G) to enable applications to utilize the network by exposing network capabilities to applications through APIs~\cite{6824752}. However, exposing critical network information to unauthorized applications will have sever security consequences~\cite{doi:10.1002/9781119293071.ch4}. Thus, communication networks using the concepts of SDN need thorough analysis of its security implications. Therefore, the third objective of this thesis revolves around security of the SDN technology and communication networks that use SDN. Beginning from investigating security challenges in SDN, this thesis evaluates existing security solutions and provides insights into strengthening the security of SDNs. The thesis also proposes novel security and mobility mechanisms for the OpenFlow implementation of SDN.

\section{Contributions of the thesis} 

The present thesis is based on three journal articles, [II, III, IV], one magazine article [VI] and two conference papers [I, V]. The author of the thesis has had the main responsibility of developing the original ideas, implementing the ideas, generating numerical results, evaluating the performance results and writing the papers [I, III, IV, VI]. The co-authors have provided constructive criticism on ideas, guided in implementation, and provided important comments on the writing. The thesis author in paper [II] was responsible for originating the idea, developing the testbed for implementing the idea, and evaluating the experimental results. The author was responsible for formulating the idea and analysing the results in paper [V]. The main contributions of this thesis are listed below:


\begin{enumerate}
	\item Architectural framework and performance evaluation of software defined cognitive networking (Paper I-II).

	\item Performance evaluation of SDN-based heterogeneous network architecture comprising different access technologies (Paper III).

	\item Analysis of security challenges in SDN, solutions proposed for those challenges and future research directions for potential security vulnerabilities in SDN (Paper IV).

	\item Analysis of security challenges and their solutions for the enabling technologies of 5G, such as SDN, cloud technologies, and virtual network functions (Paper V). 
	
	\item Performance evaluation of novel control channel security mechanism for OpenFlow (Paper VI).
\end{enumerate}

Paper [I] introduces the concept and architectural framework for SDN-based cognitive networking. The concepts of cognitive networking were mapped against the concepts of SDN. The relevant architectural components that could be integrated were studied. CRN composing the radio components of cognitive networking were sought to be integrated with the SDN controller. Hence, a testbed was developed and various experiments were carried out. The author was responsible for the creation of the idea of Software Defined Cognitive Networking, outlining the architectural framework, developing the testbed, carrying out experiments and analyzing the results from the experiments. The co-authors provided highly valuable comments and directions from the beginning until the final publication. Dr. Suneth Namal helped in developing the testbed and guided in evaluating the results. Professor Andrei suggested that I perform and validate the idea through real-time experiments. Professor Mika Ylianttila and Professor Andrei Gurtov were the supervisors.

Paper [II] continues the work started in paper [I] by extending the experimental setup with a resource allocation application, performing new experiments and shedding more light on SDN-based centralized radio resource control and allocation. The author is responsible for extending the idea of the prior paper, defining the test setup and analyzing the evaluation results. Dr. Suneth Namal was the main contributor to the work and to the paper. He helped in developing the experimental setup and analyzing the results from the experiments. Mr. Markku Jokinen and Mr. Saad Saud also helped in developing the testbed setup. Professor Andrei Gurtov provided valuable comments and suggestions.

Paper [III] proposes a centralized control framework for future heterogeneous wireless networks. The main goal of the work is to integrate and use different radio access technologies with SDN-based centralized control platform. The paper provides valuable research directions on how to enable mobility and dynamically adjust the security parameters in future wireless networks. The author is responsible for developing the idea, proposing the network architecture, performing experiments, and evaluating the results of experiments. Dr. Madhusanka Liyanage was responsible for performing the experiments of security architecture and Dr. Laszlo Bokor was responsible for performing the experiments on mobility. Professor Mika Ylianttila and Professor Andrei Gurtov were the supervisors and provided valuable suggestion and research directions.

Paper [IV] provides a thorough study of security in SDN. The paper provides an overview of SDN and security features of previous programmable networking proposals. Since SDN enables programmability and logically centralizes the network control plane, security analysis of SDN is highly important due to its vast role in future networks. The study presents security challenges in SDN by outlining the challenges associated with each of the three planes, i.e. application, control and data planes, and interfaces. Security solutions for the mentioned challenges in each plane are studied and categorized with respect to each plane and interface. The potential of SDN in strengthening network-wide security and how SDN can help improve the performance of security functions and systems such as content inspection, access control and network resilience, etc. is studied. A brief analysis of the potential of SDN for cloud and virtual networks, and the costs involved in using SDN-based security solution is provided. Various solutions for strengthening network security and the remaining loopholes according to the security dimensions of the International Telecommunication Union-Telecommunication sector (ITU-T) are presented. Future directions for addressing the potential and unresolved security challenges are also provided. The author was responsible for conducting the literature survey, selecting articles and writing the paper. Dr. Suneth Namal provided useful suggestion for improving the readability of the article. Professor Mika Ylianttila and Professor Andrei Gurtov were the supervisors.

Paper [V] describes a security framework for the control information channel between the SDN control and data planes. The paper proposes using Host Identity Protocol based solution for the control channel instead of Transport Layer Security (TLS) protocol to not only improve the security of the control channel, but also improve mobility in the OpenFlow architecture of SDN. The author was responsible for formulating the idea and analyzing the experimental results to evaluate the performance of the proposed idea. Dr. Suneth Namal was the main contributor of the work and of the paper. Professor Mika Ylianttila and Professor Andrei Gurtov were the supervisors.

Paper [VI] investigates the potential security challenges in 5G and studies the proposed solutions for those challenges. The paper also provides important insights into 5G security to instigate future research work on existing potential security and privacy vulnerabilities. Since 5G will use different technologies such as cloud computing, SDN and NFV, the paper investigates the security challenges associated with each technology and the possible solutions to those challenges. The paper also highlights the potential privacy challenges and possible solutions for those challenges. The author was responsible for conducting the literature survey, collecting the most relevant and important articles and writing the paper. All the co-authors provided contributions in text, ideas and finalizing the paper. Professor Mika Ylianttila and Professor Andrei Gurtov were the supervisors and provided important suggestions in terms of highlighting security challenges and proposed solutions.

In addition to the included original publications, the author of this thesis has participated extensively in the publication of book chapters, conference papers, magazine and journal articles. All these publications supplement the research work presented in this thesis from their own perspectives.

\section{Organization of the thesis} 

This thesis is organized as follows: in this chapter the background of the research topics, the objectives and research scope, as well as a brief summary of the contributions of this thesis are discussed. In Chapter 2, a literature review of the research topics covered in this thesis is presented. The research topics include the use of SDN in future wireless networks, and the security challenges associated with SDN and possible solutions. In Chapter 3, the main research contributions of the original research publications are summarized, the significance of the contributions is discussed, and ideas for future research are presented. This chapter consists of two parts. Part one presents the use of SDN in future wireless networks. The second part covers the security implications of using SDN in future networks, including wireless networks. In Chapter 4, some conclusions are drawn based on the research work presented in this thesis and future directions for conducting research in the area are highlighted as well.

\chapter{Literature review}

In this chapter, an overview of the literature relevant to the scope of the thesis is discussed. The chapter is divided into four parts. Section 2.1 provides an overview of SDN, whereas the use of SDN in wireless networks is presented in Section 2.2. Software defined cognitive networking is discussed in Section 2.3. The security implications of SDN and its use in future networks are described in Section 2.4.

\section{Software defined networks}

SDN~\cite{6994333} separates the network control from the data forwarding plane. The control plane is logically centralized that interacts with the forwarding plane, e.g. switches and routers through programmable APIs~\cite{Kirkpatrick:2013:SN:2500468.2500473}. By decoupling the control plane, SDN enables innovation in networking enabling programmability of the forwarding devices and the control planes. Since the control plane is softwarized and not embedded in the firmware, new control plane functions can be deployed at software speed rather than hardware or firmware product cycles. SDN proposes a three-tier architectural concept comprising the application, control and data forwarding planes, as depicted in Fig.~\ref{dr1} and described below in more detail:

\begin{figure}[h!]
\centering
\includegraphics[scale =0.60, keepaspectratio]{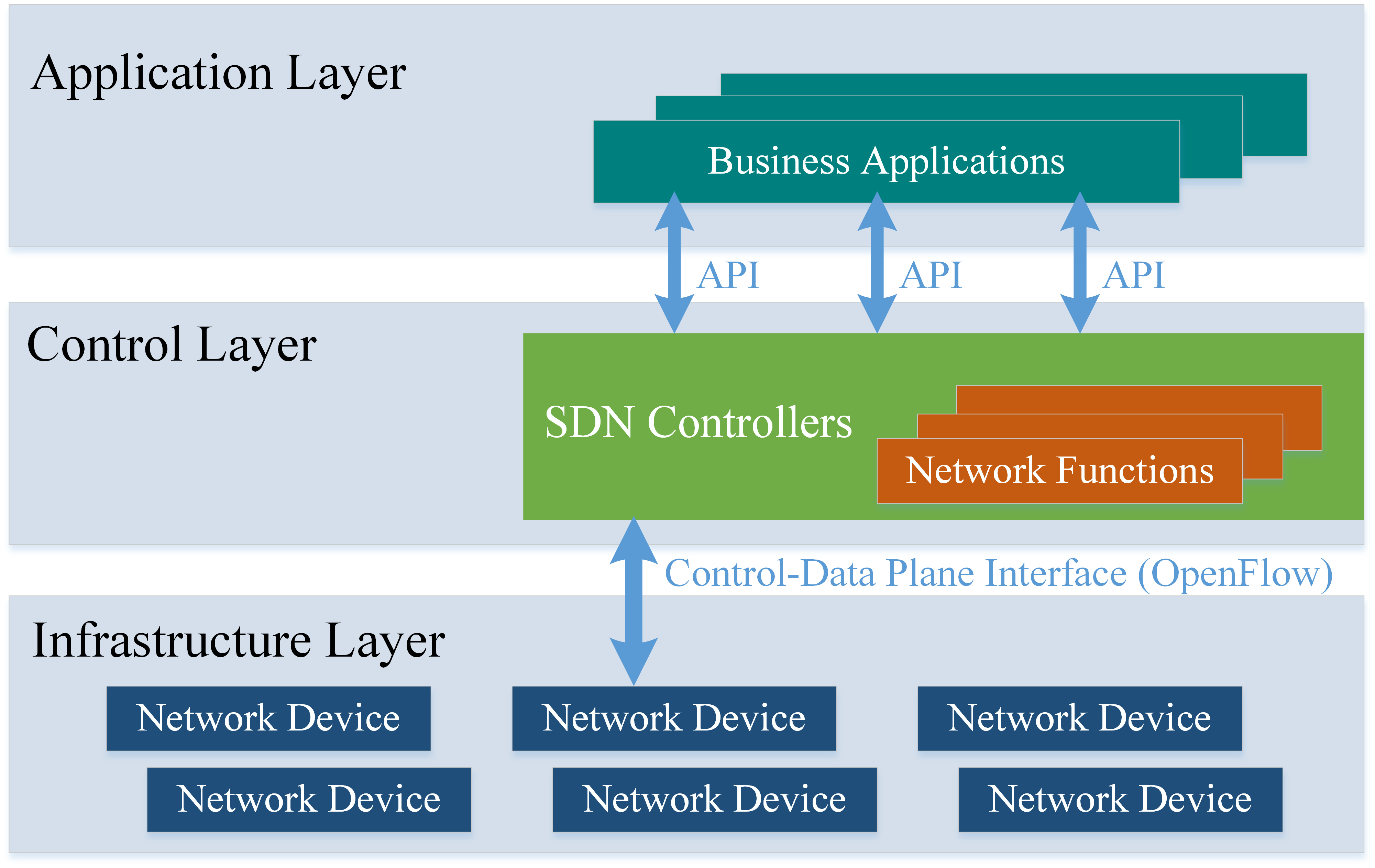}
\caption{Conceptual architecture of SDN.}
\label{dr1}
\end{figure}

\begin{itemize}
	\item \textbf{\textit{Application Plane:}} Implements the networking functions such as routing algorithms, security services and network policy implementation.
\end{itemize}

\begin{itemize}
	\item \textbf{\textit{Control Plane:}} Runs the Network Operating System (NOS), maintains the global view of the network due to its logically centralized nature, and provides hardware abstractions to applications in the application plane.
\end{itemize}

\begin{itemize}
	\item \textbf{\textit{Data Plane:}} Also called the infrastructure layer, is the combination of hardware elements used to forward traffic flows using flow forwarding instructions from the control plane.
\end{itemize}

In SDNs, most of the network functions are implemented in applications~\cite{ONF}. The SDN controller maintains a logical map of the network and hides the network complexity from applications through abstractions. The SDN controller runs the Network Operating System (NOS)~\cite{Shalimov:2013:ASS:2556610.2556621} overseeing the network resources much like a computer operating system controlling computer resources. In SDN, the controller controls the forwarding behavior of the forwarding devices such as the OpenFlow switches. The OpenFlow protocol~\cite{OFSpecification} is the most common and standardized protocol used between the controller and switches. When a flow arrives in a switch, the switch forwards the first packet(s) to the controller. The controller makes decisions on how to route the packets and installs the decisions in the switch using the OpenFlow protocol~\cite{namal2013sdn}. The decisions are in the form of flow rules describing the action on the packets of that flow. The flow rules are stored in the flow tables of switches and the controller can change the flow rules any time.

The switch flow tables also contain other information regarding flows such as packet counter values, timeout values, flags, etc. which are always visible to the controller. Using the controller, SDN applications can request flow characteristics such as timeout values or packet samples and use that to tune the network by generating flow rules. The rules, for example, can be to forward specific packets to specific ports. The decision of forwarding specific packets to specific ports might be needed for load balancing or security analysis. In the case of load balancing, the new egress ports can be connected to less loaded switches, or other forwarding devices. In the case of security, the new egress ports might be connected to security middle boxes such as firewalls or Intrusion Detection Systems (IDS). Hence, SDN enables software based innovation in communication networks. 

SDN, as a way forward for innovation in communication networks, is not only accepted by researchers and industry, but also by various standardization bodies. The Open Network Foundation (ONF)~\cite{ONF} is perceived as the leader for SDN standardization. One of the major contributions of ONF has been the OpenFlow~\cite{mckeown2008openflow} switch specification~\cite{OFSpecification}. ONF has also contributed in many other technical specifications such as the OpenFlow-Config protocol and OpenFlow testing specifications. The ITU-T started working in the SDN domain by launching two Study Groups (SGs) to develop signaling requirements and protocols, and outlining functional requirements and architectures, and a Joint Coordination Activity (JCA) to coordinate the work~\cite{ITUSDN}. ETSI started strategic collaboration with ONF for SDN support of NFV in 2014 to accelerate the adoption of open source SDN~\cite{ETSISDN}. 


Having the benefits being realized by researchers and industry alike, SDN has been sought out as the potential problem solver for next generations networks. Cellular networks, accepting the move towards 5G, are seeking SDN-based solutions to meet the growing user demands within the constraints of CapEx and OpEx. Henceforth, the thesis presents the potential of SDN in cellular networks.

\section{Software defined wireless networks}

SDN has been considered as the potential problem solver for many challenges in existing wireless communication networks~\cite{7321981},~\cite{6845049},~\cite{ahmad2015load}. Therefore, there are many proposals for using the concepts of SDN in wireless networks in general and cellular networks in particular. The use of SDN in wireless networks got attention from the OpenRoads~\cite{yap2010openroads},~\cite{yap2009stanford} platform. OpenRoads is an open-source wireless platform of OpenFlow~\cite{mckeown2008openflow}. The main idea behind OpenRoads is to enable experimentation of new ideas and solutions in production networks. The architecture of the OpenRoads consists of flow, slicing and controller that provide flexible control, virtualization and high-level abstraction. The OpenRoads architecture has worked as a foundation for many SDN-based cellular network architecture proposals.

The concepts of SDN can be used in mobile networks in different approaches which may vary from each other~\cite{7194059}. For example, two approaches that are different from each other are based on the role of SDN switches. One approach suggests maintaining the simplicity of SDN switches having mainly the forwarding functionality in the switch. All other functions such as local or global routing control decisions will be computed either in SDN controllers or SDN applications. The second approach suggests keeping some functionalities, such as packet header compression and local routing decisions, in switches or routers. Therefore, there are many approaches, each advocating its own benefits, for using SDN in cellular networks. 

A simplistic, yet promising SDN based architecture for cellular networks is presented in~\cite{6385040}. The architecture proposes four extensions to enable scalable SDN-based network architecture. First, the SDN controller applications should be capable to express policies using subscriber attributes available in subscriber information base instead of IP addresses or physical locations. Second, each switch should run a controller agent for higher scalability. Third, switches should also support more functions such as Deep Packet Inspection (DPI) and header compression. Fourth, base stations should support remote control of resources for flexible cell management. The proposal is inspired by the OpenRoads~\cite{yap2010openroads},~\cite{yap2009stanford} platform.

Mobileflow~\cite{6553677} defines a mobile network architecture called the Software Defined Mobile Network (SDMN) architecture. By decoupling the network control and data planes, the architecture is an attempt to increase the potential of innovation in mobile networks. The key elements of the system are Mobile Flow Forwarding Engine (MFFE) and Mobile Flow Controller (MFC). MFEE is a software controllable, stable and high performing user plane that has the standard mobile network tunnel processing capabilities. MFC is a logically centralized control plane that has interfaces to MFFEs, radios, and mobile network applications. The mobile network applications use the SDN-like north-bound interface and implement control planes EPC elements such as Mobility Management Entity (MME), and Policy and Charging Rules Functions (PCRF). 

The benefits of logically centralized control, that has global visibility of network state and higher control over resources, led to proposals for centralizing the core network elements of cellular networks~\cite{7429046}. Since centralization leads to scalability challenges, scalable architectures are highly important. SoftCell~\cite{Jin:2013:SSF:2535372.2535377} is a scalable core network architecture proposed to support fine-grained policy implementation for mobile devices in cellular networks. Based on the subscriber attributes and applications, SoftCell enables operators to realize high-level service policies to efficiently utilize the resources. Scalability in the core network is achieved by minimizing the state in core network through aggregation of forwarding rules. 

The concepts of SDN are also extended towards the radio part of cellular networks. For example, Software Defined Radio Access Network (SoftRAN)~\cite{Gudipati:2013:SSD:2491185.2491207} proposes a software defined centralized control plane for radio access networks (RANs). To avoid the challenges in distributed control planes, SoftRAN introduces centralized control plane for all the RANs in a defined geographical area. The control plane resides in a big BS that efficiently coordinates radio resource management among multiple BSs. Having a global view of network resources, the centralized control platform can efficiently manage interference, load, QoS, and simplify network management. Using the concept of SDN, CellSDN~\cite{li2012cellsdn} attempts to simplify the design and management of cellular data networks. CellSDN presents various approaches to design new services and prototypes with the help of open source LTE implementation. Small cell offloading through cooperative communication in SDN, to meet the growing demands of traffic, is presented in~\cite{7493644}. An overview of the benefits and research challenges in SDN-based wireless networks is presented in~\cite{7321981} and~\cite{6845049}. 

Cognitive networking~\cite{1542652} aims to automate communication networks by enabling them to make intelligent decisions. Cognitive networks understand the environment through feedback cycles and are capable to adjust the working parameters at run-time and without human interventions~\cite{6982928}. SDN has the potential to automate communication networks by enabling network programmability and relinquishing the need of per-device configurations. The logically centralized control framework provided by SDN enables applications to adjust the operation of a network according to the needs of services. Therefore, the concepts of cognitive networking and SDN are highly complementary, as described below, and thus need further investigation. Henceforth, we describe software defined cognitive networking in the following section.

\section{Software defined cognitive networking}

This section covers the importance of cognitive networking, the challenges faced by cognitive networking and presents how SDN can be a potential problem solver for those challenges.  

\subsection{Cognitive networks}

Cognitive capabilities or cognition means relating to or involving conscious intellectual activity of thinking, reasoning, or remembering~\cite{Oxford},~\cite{Webster}. Extending such capabilities to communication networks, cognitive networks have the capability to perceive the current network conditions, plan, decide and act accordingly~\cite{1542652}. The main reason cognitive capabilities are proposed for communication networks, besides others, is the lack of adaptability of communication systems to changing environments in or around the networked systems. The changing environment maybe due to changing network conditions such as congestion in some nodes, changes in services, user movements, and changes in security settings or policies.

\begin{figure}[ht]
\centering
\includegraphics[scale =0.70]{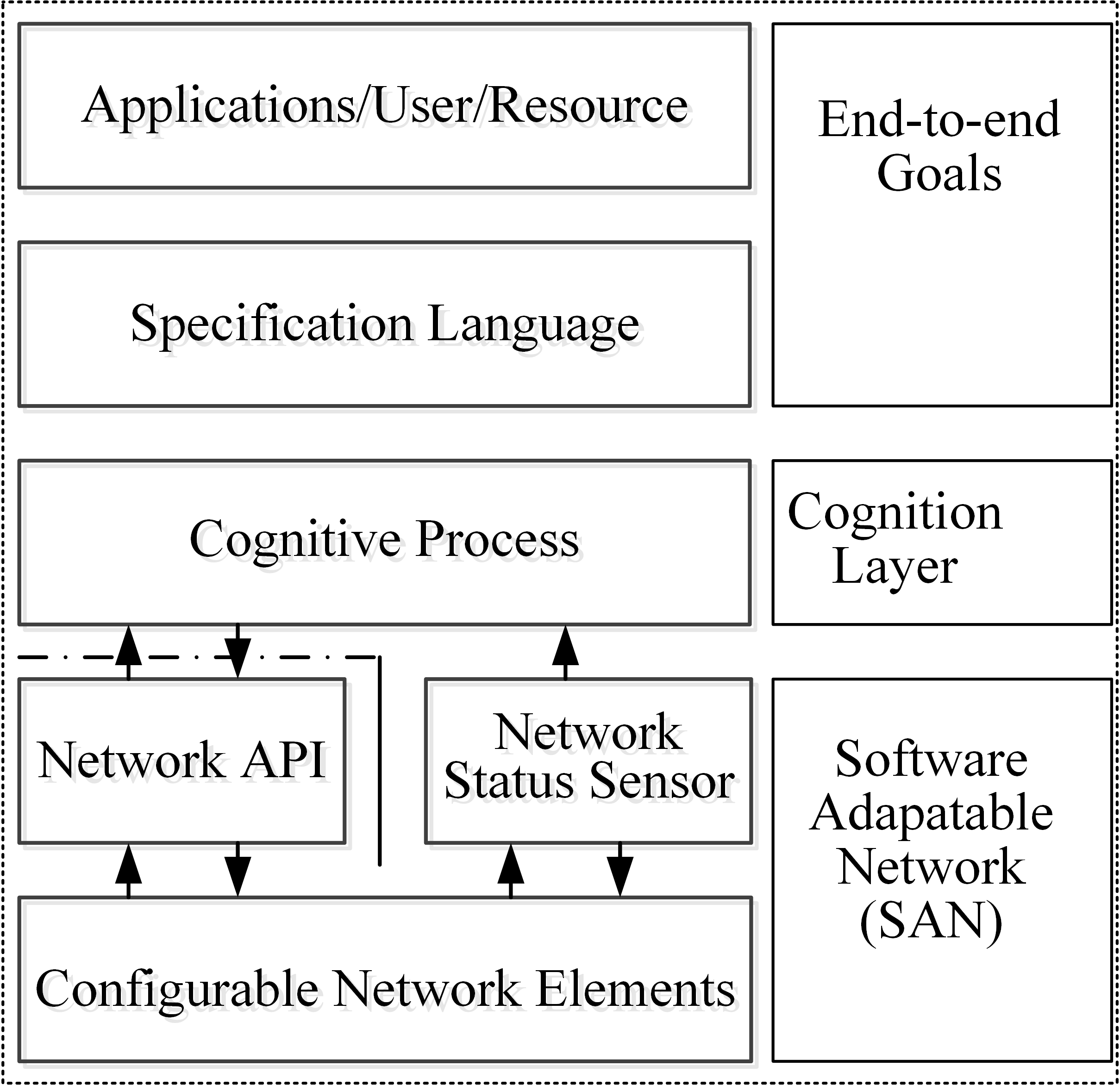}
\caption{The cognitive network framework.}
\label{dr202}
\end{figure}

A cognitive network has three layers, as presented in Fig.~\ref{dr202}. The end-to-end goals derive the behavior of the entire system and are specified by the network users, resources or applications~\cite{4050101}. The end-to-end goals are provided to the cognitive process in the cognition layer by a specification language. The cognition layer consists of the cognitive process that is responsible for the actual decision making based on the input from the end-to-end goals and the current network status. Updates from the network are provided either by network APIs or sensors. The Software Adaptable Network (SAN) layer consists of configurable network elements that can be tuned at run time by the cognitive process.

Cognitive networks aim to automate communication networks in order to respond to changes in the environment without or least possible human intervention. Cognition, in communication networks, is intended to provide better end-to-end performance; improve resource management, QoS and security; and fulfill other network goals. Efficient resource management according to the user needs and resource availability has been a major challenge for network operators. In cellular networks, the most important and scarce resource has been the frequency spectrum. Cognitive Radio (CR) systems, using cognition in radio nodes, offers a solution to the challenge by enabling spectrum sharing among users at run-time~\cite{4644051}.

A full cognitive network can be realized with the help of CRN to enable cognition at the radio or access part of the network~\cite{fortuna2009trends}. CRNs consist of primary and secondary cognitive users. Primary users are the primary license holders of the spectrum band. The secondary users use the spectrum of the primary users when it is not needed by the primary users. Both types of users use their cognitive capabilities to communicate and share the spectrum without harming each other. The main motivation behind the emergence of such cognition is the apparent lack of spectrum in the growing demands of higher bandwidth by diverse types of services~\cite{4644051}.

\subsection{Challenges faced by cognitive networks}

Cognitive networks did not realize into practical deployment because of several challenges related to the complexity of cognitive systems~\cite{fortuna2009trends},~\cite{1542652}, security challenges~\cite{6290334}~\cite{4562536}~\cite{6393525}, routing complexity~\cite{CESANA2011228} and limitations in the underlying network equipment~\cite{1542652},~\cite{7266528}. In the scope of this thesis, we focus on two challenges: first, complexity of cognitive systems, and second, limitations in the underlying network elements. 

R.W. Thomas et al.~\cite{1542652}, the main proponents of cognitive networking, describe that implementing cognitive systems requires a highly complex system in terms of architecture, operation and overhead. Thus, the designed system must outweigh such barriers with performance. The complexity is exacerbated by the limited adaptability of the underlying network elements~\cite{1542652},~\cite{Namal2016}. Cognitive networks require configurable network elements that can be tuned at run-time, as shown in Fig.~\ref{dr202}. In cognitive networks, mobility can be due to changes in the operational frequency of a node besides physical mobility of the node or UE. Similarly, CR terminals need to know the neighboring user activity, its capability, and network topology and parameters through extra control signaling while hopping between the frequency channels~\cite{AKYILDIZ20062127}. This needs extending the routing tables to include context-specific information such as frequency and propagation parameters, indicators of link quality, and end-to-end performance metrics~\cite{5577730}. 

In traditional networks, these capabilities would require strong cross-layer interaction among the vertical layers. However, interaction among the isolated layers is not only complicating the overall system, but is also highly challenging from the perspectives of overall costs of the system~\cite{1404568}. Furthermore, frequent frequency hopping will require very fast re-routing. Cross-layer designs for such frequent interactions between the physical and network layer will result in sub-optimal performance~\cite{1561928}. Moreover, with security vulnerabilities such as spoofing, the cross layer interaction might bring down the whole network~\cite{6985519}. Therefore, new networking technologies or concepts that can solve these challenges might pave the way for implementing cognitive networking. Henceforth, various proposals for SDN-based cognitive networking are discussed below.


\subsection{Software defined cognitive networks}

%

The benefits of cognitive networking can be reaped mainly when the system performance outweighs the design costs~\cite{1542652}. SDN simplifies the overall network architecture by decoupling the network control and data planes, providing infrastructure abstractions and logically centralizing the control plane. Furthermore, SDN also disintegrates the traditional layered architectures of communication networks. SDN compliments the concepts of cognitive networking in terms of adjusting or tuning the network at run-time and providing means for network automation, as depicted in Fig.~\ref{dr3}. Since SDN removes the implementation barriers of cognitive networking, the technological concepts of both must be integrated to attain the benefits of both. For example, intelligent cognitive radios coupled with adaptable OpenFlow-based network can achieve fully dynamic and automated network operations. Having said that, very limited work is done in this area and very few proposals and architectural concepts with limited applicability are available.

\begin{figure}[h]
\centering
\includegraphics[width=10cm,height=7.0cm, keepaspectratio]{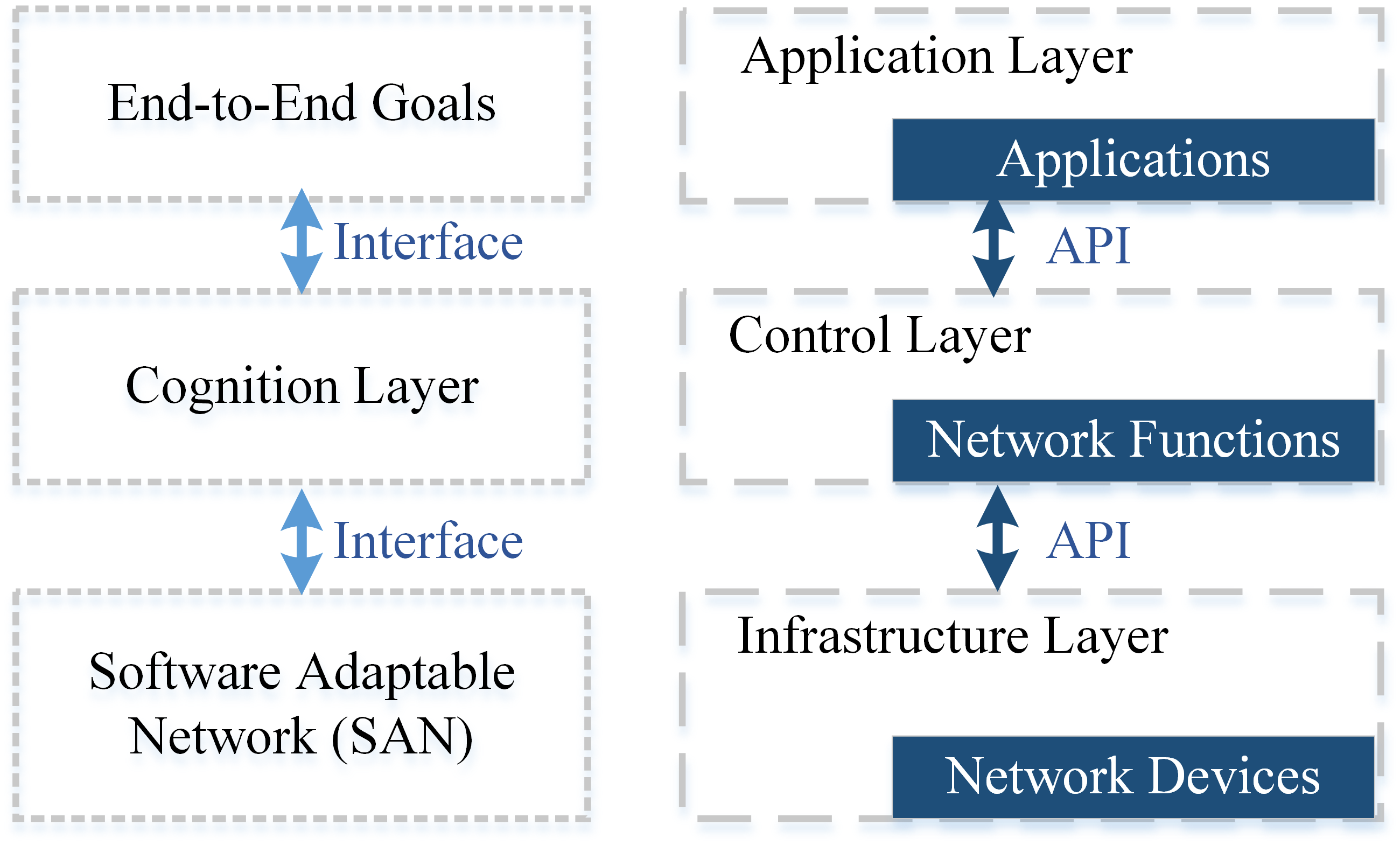}
\caption{Cognitive network and corresponding SDN architectural frameworks.}
\label{dr31}
\end{figure}

A software defined virtualization framework for CRNs using SDN is presented in~\cite{6962219}. The proposed framework reduces the control overhead in overlapping cells BSs by relegating some responsibilities to end-nodes. Similarly,~\cite{6903534} proposes a Software Defined-Cognitive Radio Network (SD-CRN) framework that enables virtualization-based resource allocation. Using multi-layer hypervisors, this framework also proposes relegating management responsibilities from CRN BSs to UEs. Both of these proposals are theoretical frameworks evaluated through simulations. The studies in~\cite{6962219} and~\cite{6903534} are very limited in scope. The focus is only on the wireless channels and interference avoidance in wireless channels. Neither of the studies mentions the architecture beyond radio nodes. 

A Joint Sub-carrier and Power Allocation (JSPA) scheme using SDN is proposed in~\cite{7174936} to reciprocally provide benefits to secondary users while cooperating with primary users. The main focus of the work is to maximize the transmission rate of the secondary user while maintaining the required QoS of primary user. This is also limited to the physical layer wireless channels, whereas modern networks require to be reconfigurable at all layers, which will be the key need of next-generation of wireless networks and services~\cite{7109098}. For example, the routing protocols and QoS parameters must be adjusted when changes in the radio layer, such as changes in operating frequency, occur.

Therefore, the work presented in this thesis aims to synchronize the dynamism of smart physical layer devices, such as cognitive radios, with the upper layers, such as backhaul devices, using SDN. The main focus of the work is increasing the overall system performance with the help of logically centralized control plane, programmable north and south-bound APIs, and configurable network elements integrated with cognitive radios. Furthermore, compared to the previous work done in this area, this thesis provides experimental evaluation of the SDN-based Heterogeneous Network (HetNet) that enables resource (spectrum) sharing, and provides mobility and security with reduced signaling overhead.

However, SDN-based architectures must be carefully designed since such architectures have their own limitations and challenges. For example, centralizing the network control and enabling programmability of networked devices might open serious security challenges besides other challenges such as scalability, delay and availability~\cite{Chen2016}. In the scope of this thesis, the security challenges and solutions in SDNs and SDN-based network architectures such as 5G are presented below.

\section{Security in software defined networks}

SDN enables innovation in communication networks by opening the network for programmability. Similarly, SDN centralizes the network control for better resource visibility, simple network management, coherent network policy enforcement and ease in new function deployment, etc. However, if overlooked, the same features can make SDNs highly vulnerable to security threats~\cite{7150550},~\cite{7226783}. For example, the centralization of the control plane makes it a favorite choice for Denial of Service (DoS)~\cite{shin2013attacking},~\cite{6212011} and saturation attacks~\cite{shin2013avant},~\cite{6733624}. Similarly, programmable networks have been prone to security vulnerabilities~\cite{casado2007ethane},~\cite{greenberg2005clean}, ~\cite{casado2006sane}. Active Networking~\cite{916839},~\cite{1243522} is one of the most prominent examples of programmable network architectures that is not used due its vulnerable nature to security threats. Indeed, security is a major challenge, and thus the security working group has been set up by ONF~\cite{6702553} to avoid such challenges in SDN.

Therefore, it is highly important to investigate the security challenges in SDNs and seek solutions for those challenges. A detailed description of security challenges in SDN and proposed solutions are provided in Paper [V], and~\cite{6702553},~\cite{7150550}. Security challenges in software defined mobile networks are highlighted in~\cite{Chen2016}. In this section, the main security challenges and proposed solutions are highlighted, mainly to convey the serious implications of security concerns related to SDNs.

\subsection{Security challenges in software defined networks}

To make it more comprehensible, the security challenges and solutions can be elaborated for each of the three SDN planes and the interfaces between them. The three planes, as already described, are the application, the control, and the data planes. The interface between the application and control planes is termed as the north-bound interface, and the one between the control and data planes is called the south-bound interface. A detailed analysis of security of SDNs is provided in Paper [IV]. In this section, the main security challenges are briefly discussed and highlighted in Table 1 with respect to the SDN planes.

\makeatletter
\newcommand{\thickhline}{%
    \noalign {\ifnum 0=`}\fi \hrule height 1pt
    \futurelet \reserved@a \@xhline
}
\begin{table*}[!ht]
\centering 
\caption{Major security threats in SDN} 
\begin{tabular}{|p{1.7 cm} |p{3 cm} |p{6.3 cm} |} 
\hline\hline 
\textbf{SDN Plane or Layer} & \textbf{Type of Threat}&\textbf{Description}  \ \\ [0.5ex] 
\thickhline 

{\textbf{Application Plane}} &{Lack of authentication \& authorization }& No compelling authentication \& authorization mechanisms for applications and more threatening in case of a large number of third-party applications.\\ \cline{2-3}

									&{Fraudulent flow rules insertion}	& Malicious or compromised applications can generate false flow rules and it is difficult to check if an application is compromised. \\ \cline{2-3}
									&{Lack of access control  \& accountability} & Difficult to implement access control \& accountability on third-party applications and nested applications that consume network resources.\\ \thickhline

{\textbf{Control Plane}} 		&{DoS attacks} &Visible nature, centralized intelligence and limited resource of the control plane are the main reasons for attracting DoS attacks.\\ \cline{2-3}
									&Unauthorized controller access & No compelling mechanisms for enforcing access control on applications.\\ \cline{2-3}
									&{Scalability \& availability}& Centralizing intelligence in one entity will most likely have scalability and availability challenges. \\ \thickhline
									
{\textbf{Data Plane}} 				&Fraudulent flow rules &Data plane is dumb and hence more susceptible to fraudulent flow rules.\\ \cline{2-3}
									&{Flooding attacks} & Flow tables of OpenFlow switches can store a finite or limited number of flow rules. \\ \cline{2-3}
									&{Controller hijacking or compromise}& Data Plane is solely dependent on the control plane that makes the data plane security dependent on controller security. \\ \cline{2-3}
									&TCP-Level attacks &TLS is susceptible to TCP-level attacks.\\ \cline{2-3}
									&Man-in-the middle attack &This is due to optional use of TLS, and complexity in configuration of TLS.\\ \thickhline

\end{tabular}
\label{table:attk} 
\end{table*}

\subsubsection{Application plane security challenges}
As described earlier, applications implement most of the network functionality without being tied to the network. Therefore, security of applications has big implications in SDNs. Applications must be authenticated and authorized before generating flow rules. However, there are no established authorization mechanisms between the controller and applications, as demonstrated in~\cite{Lee:2016:SSS:2876019.2876024}. Different applications have different privileges, thus, a security model must be put to isolate applications~\cite{6553676}. In this case also, there are no compelling mechanisms to provide differentiated access to applications based on their privileges. The challenge is further complicated by nested applications~\cite{tsou2012use} in which keeping track of a malicious application in a legitimate application is highly challenging. Furthermore, no mechanisms are defined for accountability of nested applications~\cite{hart2013sdnsec}.

\subsubsection{Control plane security challenges}
The SDN controller must authenticate applications before providing network stats to applications. The controller, however, is facing many security challenges itself. Centralizing the network control into the controller means a huge number of requests from applications and the underlying controlled devices. This can make the controller a potential bottleneck due to scalability or resource limitations~\cite{6553676}. This scalability limitation will open doors for DoS attacks. A network fingerprinting method has been used in~\cite{Shin:2013:ASN:2491185.2491220} to identify a network as an SDN and launch attack. Knowing that the switch has to send the flow setup request to the controller for each new flow, a DoS attack can be easily targeted towards the controller. To sum up, the controller is the most crucial part of SDN, on the one hand, and the most vulnerable to security attacks such as DoS and Distributed DoS (DDoS) attacks, on the other hand.

\subsubsection{Data plane security challenges}
The SDN data plane, such as the OpenFlow switches, stores flows after sending the first packet to the controller for setting up the flow rules. The flow tables that maintain the flow rules, and the buffer that store the unsolicited flows, have physical capacity limitations. These limitations can also be exploited to launch a DoS attack against the OpenFlow switches, as demonstrated in~\cite{Shin:2013:ASN:2491185.2491220}. Furthermore, the forwarding devices in SDN are rendered simple and are highly dependent on the controller. This means that the forwarding devices have no capability to differentiate between genuine and flawed or malicious flow rules. Similarly, if the controller is compromised, failed, or the link to the controller is broken, the data plane will be naturally compromised or will become non-responsive to newly arriving flows. 

\subsubsection{Interfaces security challenges}

The OpenFlow switch specification~\cite{heller2009openflow} suggests Transport Layer Security (TLS)~\cite{dierks2008transport} and Datagram Transport Layer Security (DTLS)~\cite{rescorla2012datagram} for the south-bound interface, i.e. the OpenFlow protocol. However, the standard of TLS is not specified and the security feature is left optional~\cite{7150550}. In the OpenFlow vulnerability assessment study~\cite{Benton:2013:OVA:2491185.2491222}, it is described that the optional use of TLS leaves the control channel between controllers and switches open to security threats such as man-in-the-middle attack and fraudulent rule insertions. Furthermore, the configuration of TLS is highly complex, making it a technical barrier for operators to use it~\cite{Benton:2013:OVA:2491185.2491222}. Hence, security of the south-bound API in SDN is still an open research challenge. The north-bound interface, on the other hand, has no explicitly defined security architectures. More specifically, communication between remote applications and the controller is not even properly investigated. Hence, this is also an open research challenge.

\subsection{Security solutions for software defined networks}

By centralizing the network intelligence and enabling network programmability, SDN facilitates quick identification and remediation of security threats. This interesting feature of SDN, the global network stats visibility and control, supports run-time security monitoring, analysis and response through a cycle of harvesting intelligence from the data forwarding elements using programmable APIs. In contrast to the distributed control in network elements requiring perimeter-based expensive security systems that often have contradictory security policies, SDN enables coherent network-wide security policy enforcement. With software-based systems, SDN brings forth the idea of software-defined security to experiment and innovate novel security systems, as presented in~\cite{shin2013fresco}. As an example, FLOWGUARD~\cite{Hu:2014:FBR:2620728.2620749} is a comprehensive software firewall framework for OpenFlow networks that ensures consistent firewall policy implementation throughout the network.

The OpenFlow SDN architecture offers control over the forwarding elements from centralized control point where security services insertion is simplified. For example, security applications can request the packet samples through the controller using simple sampling mechanisms such as FleXam~\cite{Shirali-Shahreza:2013:FFS:2491185.2491215} that sends full or part of packets to the control based on counter values in the switch. An application can perform analysis and then through the controller can change the flow rules such as drop consecutive packets of a particular flow due to suspicion based on security analysis of the application. Therefore, a number of proposals have been put forward for security of SDNs and securing communication networks while leveraging the concepts of SDN. Below we describe the security solutions with respect to the SDN planes and present some of the major platforms and proposals in Table 2. The platforms outlined in Table 2 are discussed in Paper [VI] in detail.

\begin{table*}[!ht]
\centering 
\caption{Security solutions for SDN} 
\begin{tabular} {|p{1.7 cm} |p{2.8 cm} |p{3.2 cm} |p{3.8 cm} |}
\hline\hline 
\textbf{SDN Plane} & \textbf{Security Solution }&\textbf{Targeted Threat} &\textbf{Solution Type} \ \\ [0.5ex] 
\thickhline 

{\textbf{Application Plane}} &FRESCO~\cite{shin2013fresco} & Threats within/from applications & Security applications development framework \\ \cline{2-4}
	
									&PermOF ~\cite{wen2013towards}	 & Access control &Applications permission system \\ \cline{2-4}
									&Assertion~\cite{beckett2014assertion}			& Flow rules contradiction &Applications debugging framework \\ \cline{2-4}
									&Flover ~\cite{6654813} & Security policy violation & Security policy verification application \\ \cline{2-4}
									&OFTesting ~\cite{canini2011automating}			& Faulty OF programs  & Applications testing framework \\ \thickhline
	
{\textbf{Control Plane}} 		&SE-Floodlight ~\cite{SEFloodLight}	&Applications authorization&Secure controller architecture \& secure App-Ctrl API\\ \cline{2-4}
									&Hybrid Ctrl ~\cite{6531863} & Controller scalability & Hybrid (reactive/proactive) controller architecture\\ \cline{2-4}
									&DISCO ~\cite{6838330}, ~\cite{6838273}& Controller scalability &Distributed controller architecture\\ \cline{2-4}
									&Ctrl-Placement ~\cite{heller2012controller}, ~\cite{6821736}, ~\cite{6573050}&Controller availability& Controller placement frameworks\\ \cline{2-4}
									&HyperFlow ~\cite{tootoonchian2010hyperflow}&Controller availability&Distributed control plane \\ \cline{2-4}
									&DoSDetection ~\cite{braga2010lightweight}&DDoS attack & Detection framework \\ \thickhline
{\textbf{Data Plane}} 				&FortNOX~\cite{Porras:2012:SEK:2342441.2342466} &Flow rules contradictions & Controller framework\\ \cline{2-4}
									&FlowChecker ~\cite{al2010flowchecker} & Faulty flow rules & Configuration verification tool\\ \cline{2-4}
									&VeriFlow ~\cite{Khurshid:2012:VVN:2377677.2377766}& Faulty flow rules & Network debugging tool \\ \cline{2-4}
									&Resonance ~\cite{nayak2009resonance}& Access control& Access control \& policy enforcement framework\\ \cline{2-4}
									&CPRecovery ~\cite{6212011} & Controller availability & Controller replication framework \\ \thickhline

\end{tabular}
\label{table:planes} 
\end{table*}

\subsubsection{Security solutions for applications}

In SDN, mechanisms that authenticate and authorize applications, and verify the flow rules generated by such applications are highly important. OperationCheckpoint~\cite{6980437} and PermOF~\cite{wen2013towards} are systems that allocate permissions to applications and set limits on the operations of applications. The proposed mechanisms authorize applications for specific actions and thus can prevent non-authorized modifications of flow rules. One of the main benefits of the proposed system in OperationCheckpoint~\cite{6980437} is to secure the north-bound interface, thus, delimiting applications to work in the defined jurisdiction. An assertion language that supports verifying and debugging SDN applications is proposed in~\cite{beckett2014assertion}. The mechanisms proposed in~\cite{beckett2014assertion} also enable verifying flow rules dynamically as the rules are produced by applications. For developing security applications, FRESCO~\cite{shin2013fresco} provides a framework to rapidly develop and deploy OpenFlow security applications.

\subsubsection{Security solutions for controllers}

Since the controller plays a crucial role in SDNs, there are many proposals and solutions that strengthen the controller security. The requirements of robust, secure and resilient SDN controllers are presented in~\cite{7258233}, alongside the analysis of the gap between security level of current controllers and potential security solutions. Security of the SDN controller is, in fact, multi-folded requiring security solutions for both interfaces, solutions to mitigate saturation attacks through increased scalability, DoS or DDoS attack-specific solutions and reliable controller placement. Security-enhanced (SE) Floodlight controller~\cite{SEFloodLight} is a secure version of the OpenFlow Floodlight controller~\cite{FloodLight}. SE Floodlight controller secures the northbound API, authenticates and authorizes applications and verifies flow rules. There are also northbound API-specific security solutions such as OperationCheckpoint~\cite{6980437}, and mechanisms to increase trust relationship between controllers and applications, as demonstrated in~\cite{7116153}. FortNOX~\cite{Porras:2012:SEK:2342441.2342466} improves the NOX controller~\cite{gude2008nox} to avoid contradictions in flow rules generated by applications.

To avoid the single point of failure problem, distributed yet logically centralized controllers are proposed. For example, HyperFlow~\cite{tootoonchian2010hyperflow} proposes a scalable event-based multi-controller architecture. Multiple distributed controllers, being logically centralized, take local decisions to minimize latency in flow setup. Increasing scalability through increasing processing capabilities of the controllers is another approach to avoid saturation attacks~\cite{Voellmy:2012:SSD:2342356.2342414}. To secure the controller from being fingerprinted for saturation or DoS attacks, AVANT-GUARD~\cite{Shin:2013:ASV:2508859.2516684} introduces connection migration for the data plane to remove failed TCP sessions and thus reduce the number of data plane to control plane interactions. For reliable controller placement, there is no one-fits-all way~\cite{Heller:2012:CPP:2342441.2342444}, but a tradeoff among desired goals such as order of redundancy vs reliability and reliability vs. latency~\cite{6821736}.

\subsubsection{Security solutions for data plane}

The data plane security solutions range from security of the interfaces, and flow tables to network planning and segmentation. Standardizing and mandating the security proposals such as the versions of TLS and DTLS are the potential solutions to the interface or link layer security. TLS, when properly configured, can provide privacy and data integrity between two communication parties such as the controller and data forwarding elements. Furthermore, cryptographic security protocols such as Host Identity Protocol (HIP)-based solutions~\cite{5451761},~\cite{7980675} can ensure both payload and control information security~\cite{6702540}. 

The security of flow tables can be ensured by flow rules verification mechanisms as described in~\cite{Porras:2012:SEK:2342441.2342466},~\cite{Khurshid:2012:VVN:2377677.2377766},~\cite{Mai:2011:DDP:2043164.2018470} besides authentication and authorization mechanisms for applications that generate flow rules, as described in the previous sections. Proper network planning and segmentation is necessary to avoid over-provisioning of the control plane that can cause controller saturation rendering the data plane non-responsive to network traffic and new flow arrivals. Resilience mechanisms enabling the network to operate in case of controller failure~\cite{6532493}, controller-switch link failure~\cite{6838393}, and disruption-free controller replacement~\cite{Vanbever:2013:HCE:2491185.2491194} are proposed to keep the data plane forwarding intact under such circumstances.

\subsection{Using SDN to improve security of communication networks}

SDN simplifies both development and deployment of novel security systems due to the programmable nature of its network components and centralized control of traffic traversing the network~\cite{6778942}. By providing packet level information access to the controller and applications on top of it, SDN enables real-time traffic monitoring and response to abnormal traffic behavior. Packet-level traffic analysis has been challenging in traditional networks due to large volume of traffic, for example in large Data Center Networks (DCNs), and the complexity in tracking multiple network components~\cite{Zhu:2015:PTL:2829988.2787483}. Indeed, SDN brings an opportunity for such granularity, yet the centralized control plane needs novel solutions to avoid congestion. Therefore, scalable per-flow sampling mechanisms~\cite{6649444} are proposed to operate at line-rate in the data plane while maintaining fair load on the controller.

The simplicity of implementation and deployment of new services such as Multi-protocol Label Switching (MPLS) Virtual Private Networks (MPLS VPNs) is demonstrated in~\cite{sharafat2011mpls}. The benefits of centralized control for secure cloud computing, and cellular networks are described in~\cite{hao2010secure},~\cite{okwuibe2018cloud} and~\cite{7373252},~\cite{DING201494},~\cite{8088621},~\cite{7399014}, respectively. Benefiting from the concepts of SDN for improving anomaly detection systems in small and home networks is demonstrated in~\cite{Mehdi2011}. The OpenSafe~\cite{ballard2010extensible} system using A Language for Arbitrary Route Management for Security (ALARMS) to route traffic to monitoring systems is an example of economical deployment of security monitoring systems in SDN. The adequacy of SDN has also been proved for deploying security middle boxes without changing the network architecture or modification in middle boxes, as demonstrated in~\cite{Qazi:2013:SMP:2534169.2486022}. Using SDN,~\cite{bates2014let} explores new opportunities to deal with the challenges in network forensics systems. The solution proposes Provenance Verification Point component to observe covert communication between compromised nodes. The interesting theme of the work is using the network itself as an observer, leveraging the centralized SDN control plane obtaining information from switches distributed in the network.

\subsection{Open source security}

ONF has published three security Technical recommendations; i) Security Foundation Requirements for SDN controllers, ii) Threat Analysis for the SDN Architecture, and iii) Principles and Practices for Securing Software-Defined Networks~\cite{ONFSecurity}. On the development side, there are many open source security development projects alongside the open source SDN development projects. The OpenFlowSec~\cite{OpenFlowSec} consortium has built the SDN security suite by extending the Floodlight~\cite{FloodLight} OpenFlow controller. The software suite includes a security actuator to invoke refined security logic, and a Bothunter to perform passive security analysis in OpenFlow networks. The OpenDayLight foundation~\cite{OpenDayLight} initiated the project AAA (Authentication, Authorization and Accounting)~\cite{OpenDayLightAAA} to develop security modules to authenticate identities, authorize access, and maintain the records of access to resources.

\chapter{Summary of the original articles}

This chapter briefly summarizes the research work published in the journal articles and conference proceedings. The work has been divided into two sections. Section 3.1 describes using SDN for integrating diverse RATs, enabling intelligent resource sharing, and providing cost efficient mobility and security parameters adjustments in future networks. Section 3.2 describes the security challenges in SDN and the opportunities in terms of security in using SDN in future networks.

\section{SDN-based cognitive and heterogeneous networking}

\begin{figure}[h!]
\centering
\includegraphics[width=8cm,height=5.0cm, keepaspectratio]{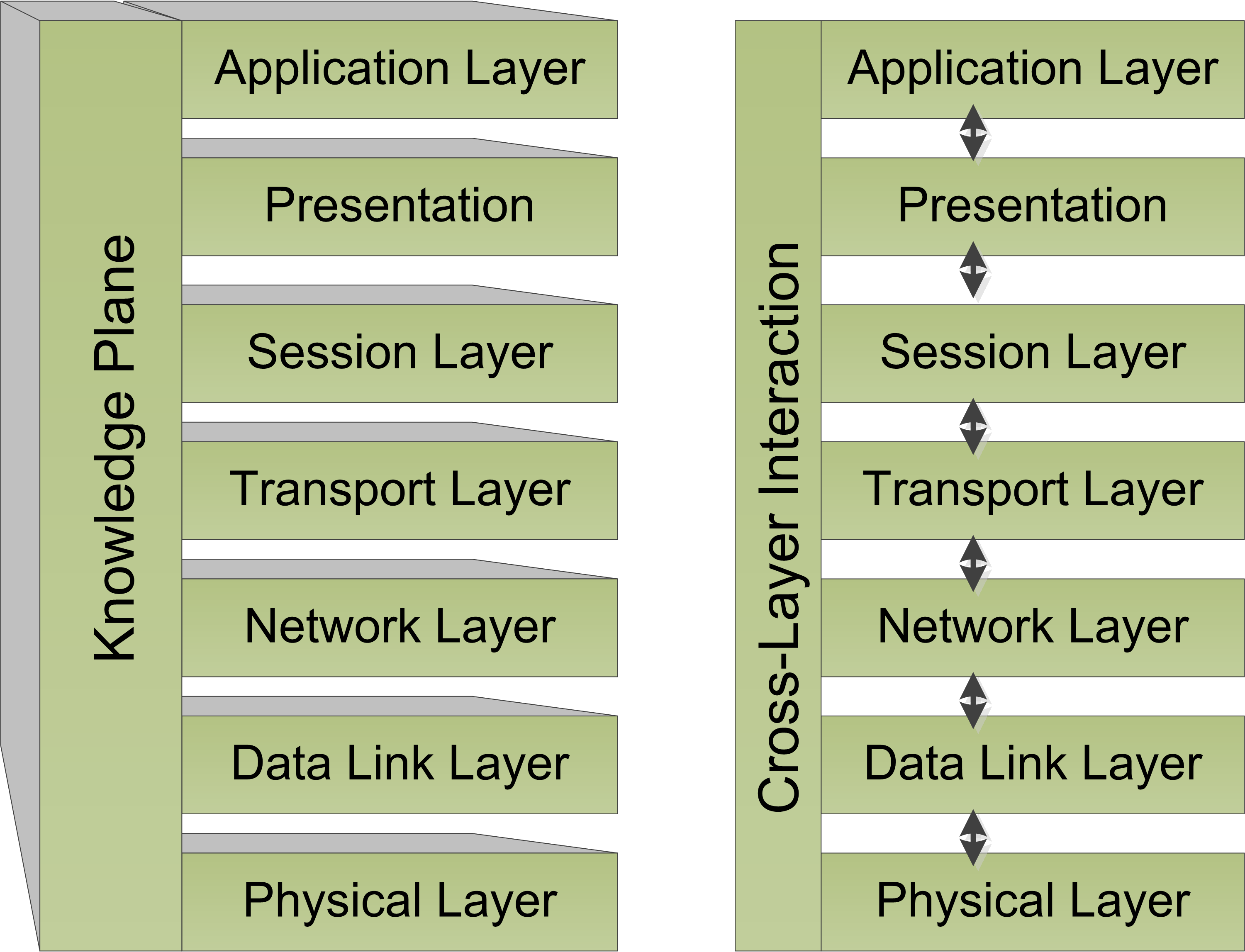}
\caption{The knowledge plane and cross-layer interaction among the layers.}
\label{dr2}
\end{figure}

Papers [I, II, III] discuss the need of dynamic network systems and propose futuristic network architectures using the concepts of SDN. Using SDN, a HetNet architecture has been the main focus that use cognitive networking and multiple access technologies to meet the demands of future services. Cognitive networks sense the current state and context, adapt to contextual changes and apply control loop systems to learn and update itself for future actions without human intervention~\cite{thomas2005cogn}. Such systems require cross-layer interactions, as shown in the right-hand side of Fig.~\ref{dr2}. Therefore, the idea of Knowledge Plane (KP), spanning across all layers, as shown in the left-hand side of Fig.~\ref{dr2}, is proposed by Clark et al.~\cite{clark2003knowledge}. The main idea of the KP proposal is to make the network self and surrounding-aware, capable to learn, make decisions and act on those decisions~\cite{fortuna2009trends}. 

\begin{figure}[h]
\centering
\includegraphics[width=9cm,height=6.0cm, keepaspectratio]{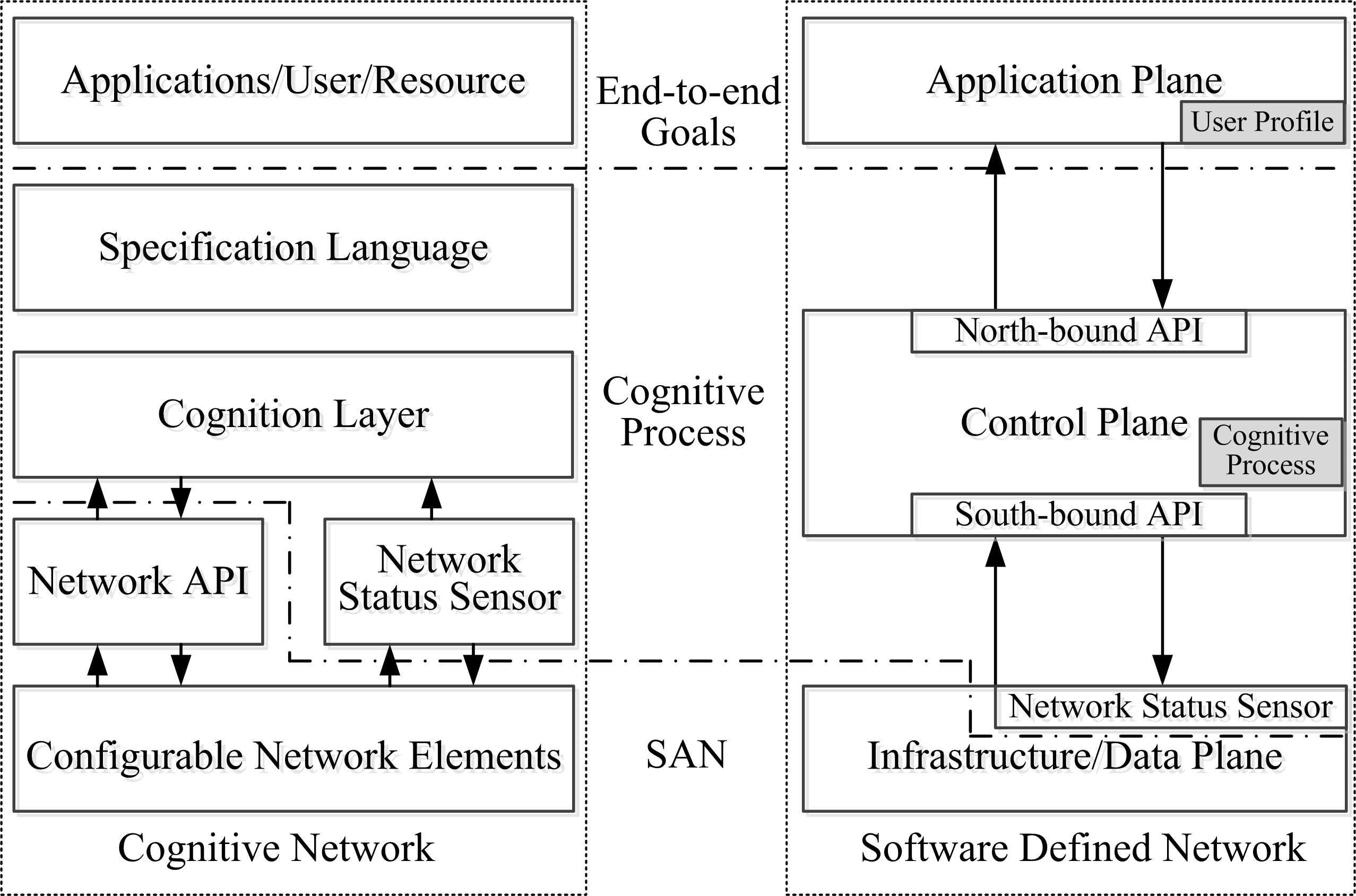}
\caption{Cognitive networking and corresponding SDN architectural frameworks.}
\label{dr3}
\end{figure}

However, strict isolation between the vertical layers made it difficult to realize KP like architectures or cross layer interaction. For example, the cross-layer interaction requires the network to modify one or several layers of the network stack in its member nodes at run-time. This and other challenges such as manual configurations, as described in Section 3.2, are already solved by SDN. A simplified architectural mapping of the concepts of both technologies is presented in Fig.~\ref{dr3}. Therefore, Paper [I] investigates the possibility of integrating the concepts of SDN and cognitive networking, and provides the initial results. Paper [II] extends the concepts by developing a fully automated SDN-based cognitive network. Furthermore, Paper [III] focuses on using SDN for enabling a heterogeneous network architecture that provides mobility between different RATs, dynamically adjusts security parameters, as well as dynamically shares spectrum among multiple user nodes.

\subsection{Towards software defined cognitive networking}

Paper [I] proposed the idea of combining the concepts of SDN and cognitive networking for efficient and dynamic use of resources (e.g. spectrum). Cognitive Radio (CR) implements the physical layer part of the cognitive networks, making a CRN capable to sense a free or unused spectrum and use that during the time it is available (more details about CR in Papers [II, III]). By removing the layering intricacies and centralizing decision making, SDN makes the idea of a whole cognitive network realistic. The architectural concepts of cognitive networking and SDN are mapped, as shown in Fig.~\ref{dr3}. Both concepts facilitate network automation, albeit the differences from the implementation point of view. Therefore, the outcome has been of great interest.

\begin{figure}[h]
\centering
\includegraphics[width=12cm,height=9.0cm, keepaspectratio]{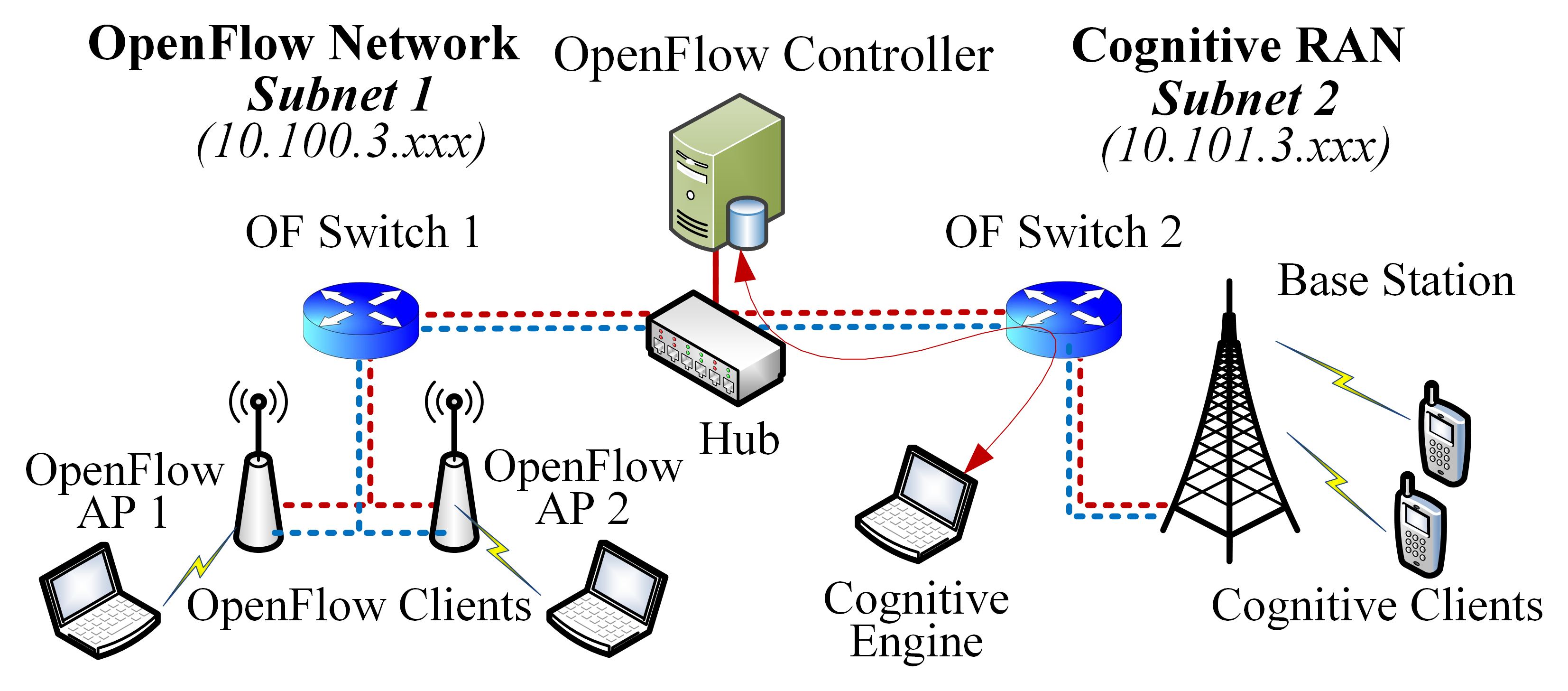}
\caption{SDN-based cognitive radio network test setup.}
\label{dr4}
\end{figure}

A testbed was developed that enabled the control of CRs through the SDN controller. The three main parts of the testbed were a CRN, OpenFlow wireless network, and the Floodlight OpenFlow controller-based SDN control plane. The cognitive engine, working as the spectrum sharing decision making entity, was attached to the SDN controller and the base stations. The details of the testbed (Fig.~\ref{dr4}) implementation are given in Paper [I]. The main aim of this initial research was to evaluate the potential of SDN in controlling the radio resources using CRs. The experimental results revealed the performance improvement in terms of throughput, QoS, and delay, as presented in Paper [I].

\begin{figure}[h]
\centering
\includegraphics[scale =0.25]{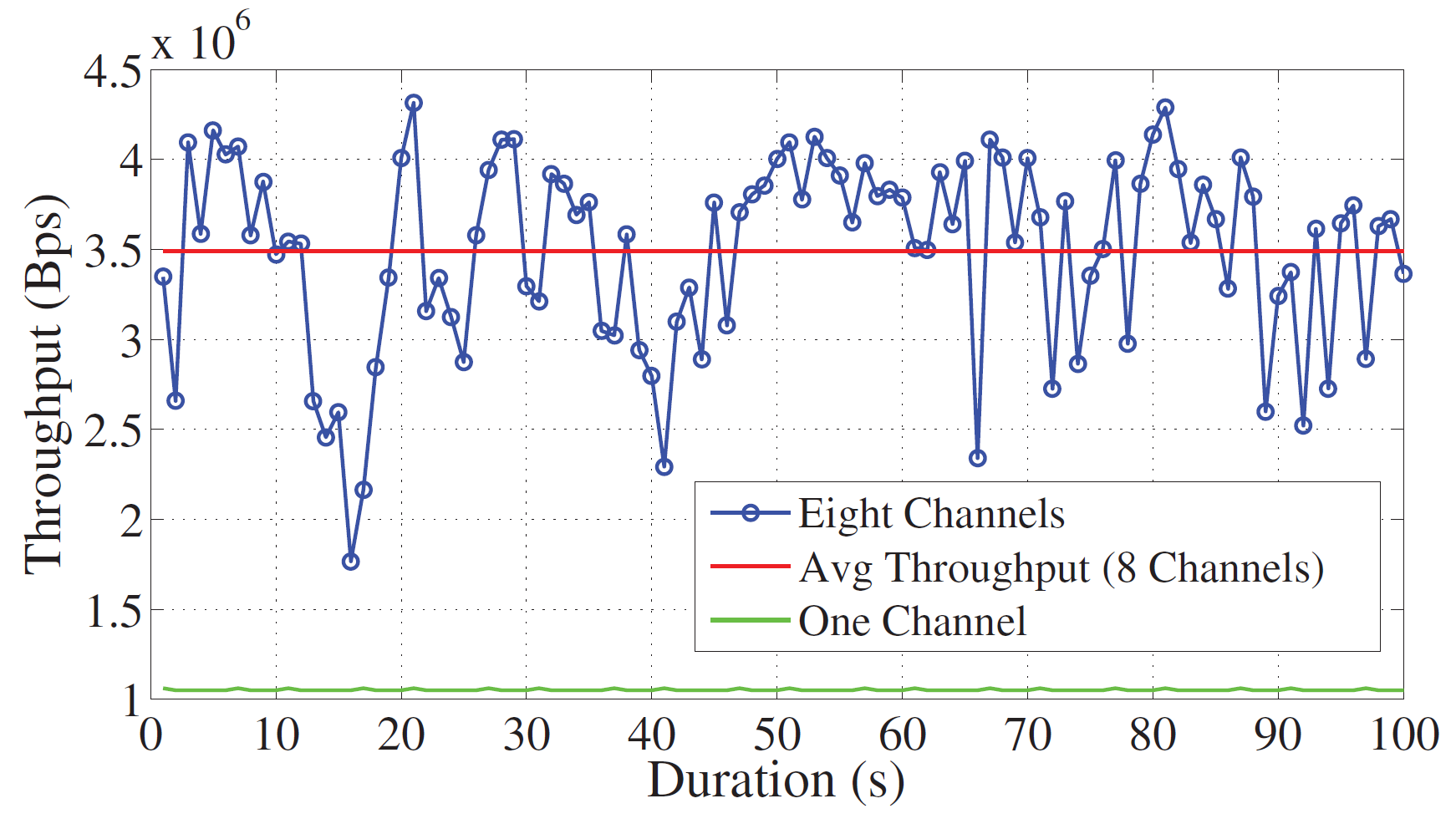}
\caption{TCP throughput in a cognitive environment.}
\label{dr41}
\end{figure}

\begin{figure}[h]
\centering
\includegraphics[scale =0.20]{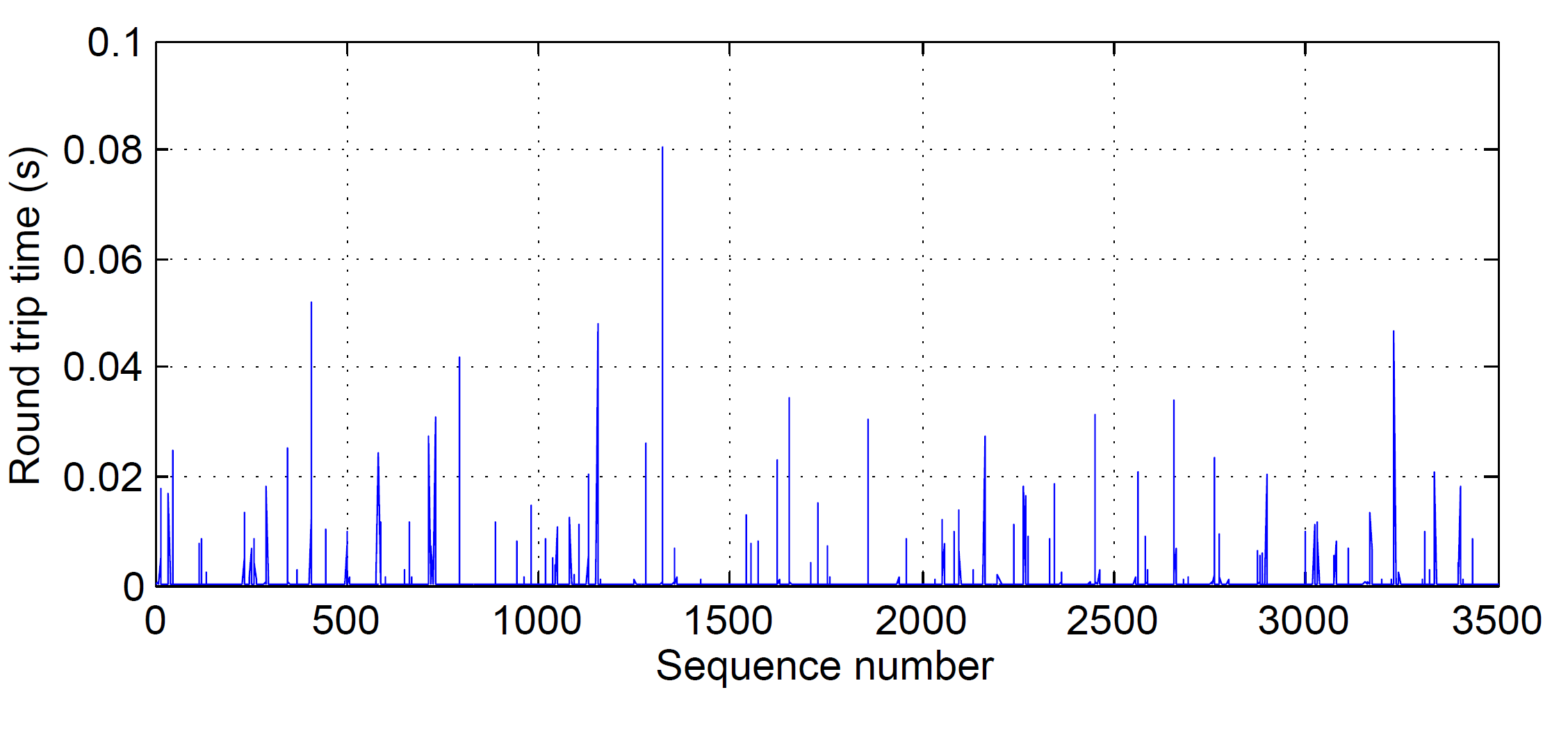}
\caption{Round-trip time between OpenFlow and cognitive clients.}
\label{dr42}
\end{figure}

To provide an overview of the performance improvement, the throughput measurement results are depicted in Fig.~\ref{dr41}. The measurements are taken between the OpenFlow and cognitive clients. TCP traffic was generated between the clients on one and eight frequency channels. Using a single frequency channel, the average throughput remained 1.15 MBps, whereas using cognition among the eight frequency channels, the average throughput increased to 3.5 MBps. Similarly, the average round-trip-time (RTT) for TCP packets and the corresponding ACK (Acknowledgment) packets between the clients remained well below 20ms, as shown in Fig.~\ref{dr42}.

\subsection{Implementation of full OpenFlow based CRN}

Paper [II] describes the implementation of OpenFlow based CRN architecture. In this work, we realized that the end-to-end goals defined in cognitive networking can be implemented in the application plane or the network management plane. Those end-to-end goals can be deployed through the control plane in the underlying network using the north and south-bound APIs. The cognitive process can be implemented as an application or as a software module in the control plane. SAN elements can be realized as OpenFlow switches that are tunable at run-time by the controller using the south-bound API, for example, the OpenFlow protocol. By sensing the free spectrum, cognitive radios can provide the sensed information to the cognitive process using the same API.

\begin{figure}[h]
\centering
\includegraphics[scale =0.20]{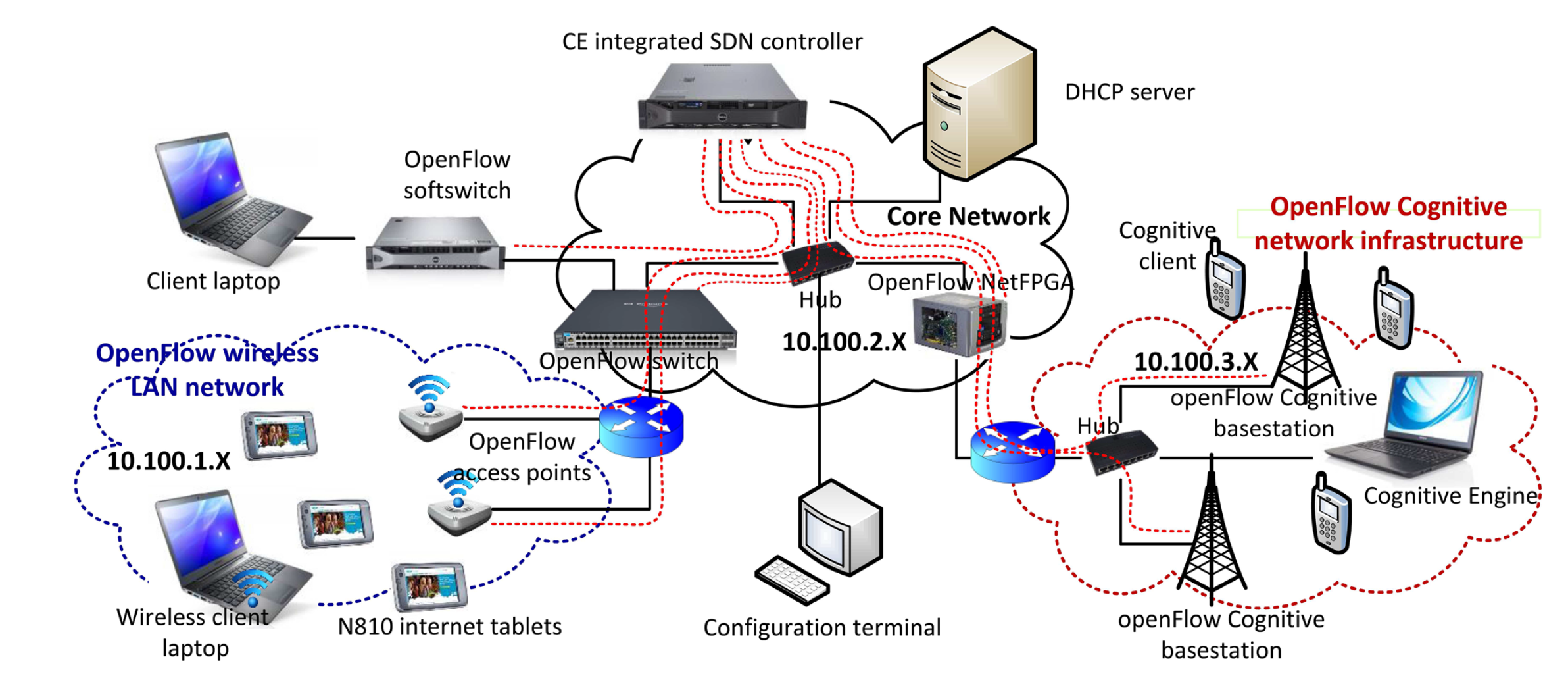}
\caption{Extended testbed implementation for OpenFlow-based CRN.}
\label{dr43}
\end{figure}

Therefore, the work presented in Paper [II] integrates software defined radio (SDR)-based CRs to SDN-based control plane. The aim is to synchronize the dynamism of SDN-based control platforms with the dynamism of cognitive radios for the efficient use of resources (spectrum in our case). Hence, the testbed was extended that integrated the concepts and technological components of SDNs and cognitive networks, as presented in Fig.~\ref{dr43}. In the testbed, the Base Station (BS) was modified to work in a fashion similar to an OpenFlow switch, e.g. maintain a flow table which can be updated or modified by the OpenFlow controller. A Cognitive Engine (CE) working as an OpenFlow application is integrated into the SDN controller. The controller installs the flow rules based on the decisions from the CE.

\begin{figure}[h]
\centering
\includegraphics[scale =0.40]{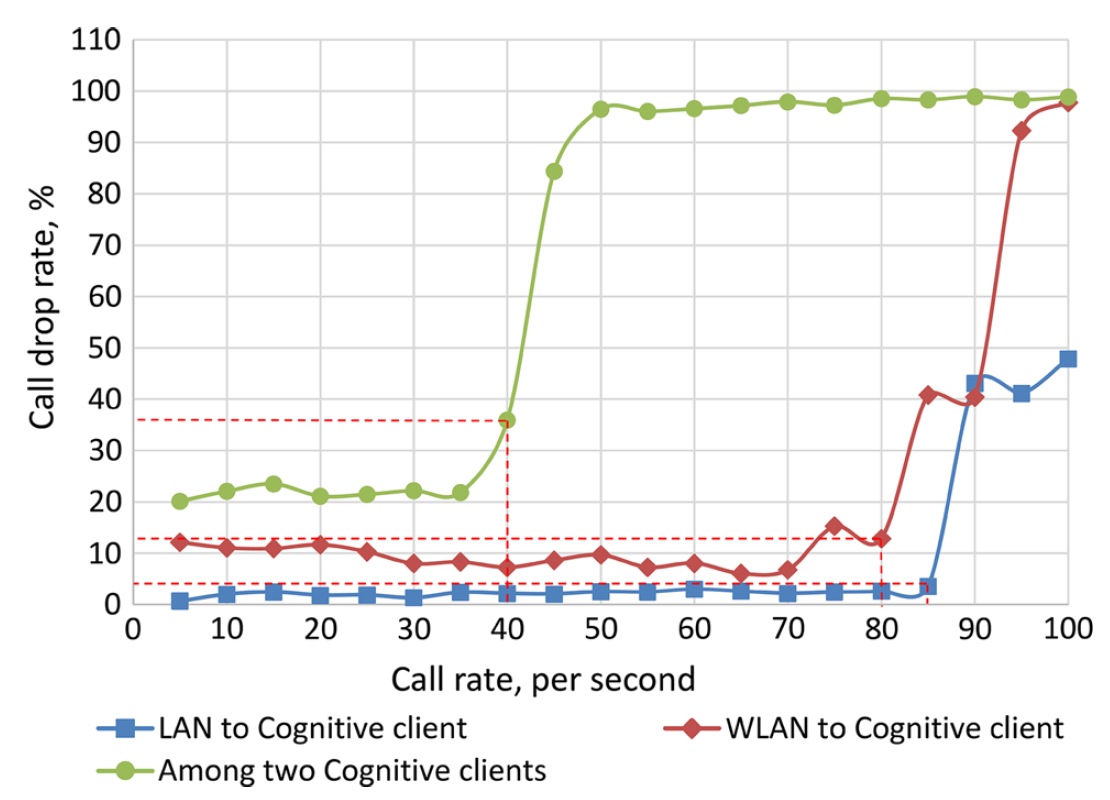}
\caption{Network performance results in different scenarios.}
\label{dr44}
\end{figure}

The performance of the proposed architecture was evaluated in terms of latency in flow setup, QoS for VoIP calls, and measurement of response times between different network segments. For brevity, I will describe the VoIP performance in terms of latency and call rate per second that highlights the number of successful calls per second in different scenarios. Detailed evaluation of the experimental results is presented in Paper [II]. The latency requirement for VoIP call is 150 ms, although 151 ms to 300 ms might be an acceptable one-way delay~~\cite{Jiang:2016:VII:2934872.2934907}. Therefore, we consider 150 ms RTT between the two clients as the threshold and calls with delays above the threshold are dropped. As shown in Fig.~\ref{dr44}, the call drop rate is very low when the resources are not shared among two cognitive clients. Though the call drop rate increases when two cognitive clients share the resources, however, our main aim is sharing resources dynamically among the clients using SDN-based centralized control framework.

\subsection{Enabling heterogeneity in cellular networks using SDN}

\begin{figure}[h]
\centering
\includegraphics[width=12cm,height=11.0cm, keepaspectratio]{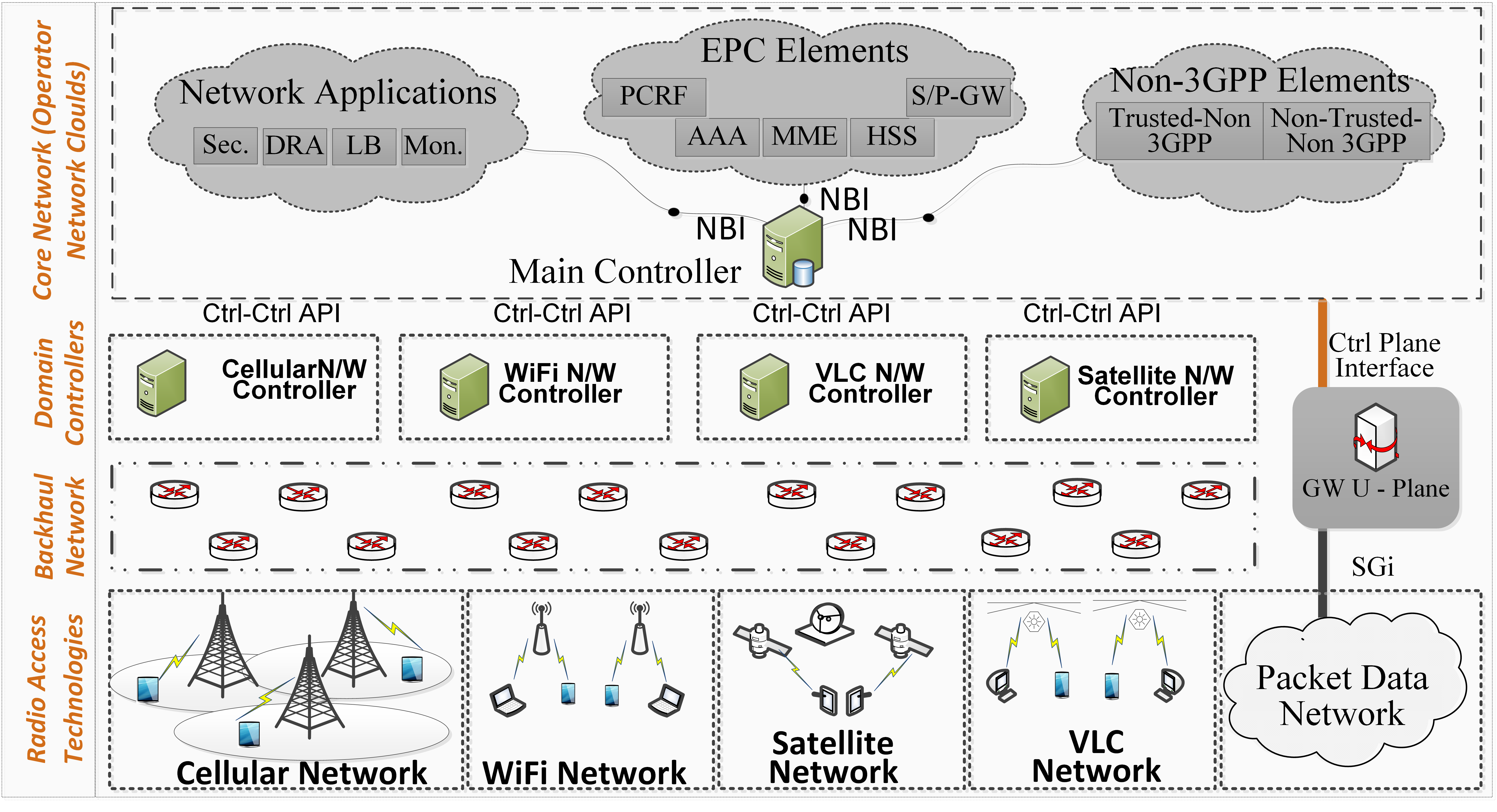}
\caption{SDN-based heterogeneous network architecture.}
\label{dr5}
\end{figure}

Paper [III] presents a HetNet architecture leveraging SDN. Next generations of mobile networks, such as 5G, will need to use multiple RATs together to extend the connectivity to existing and newly invented digital devices~\cite{6812298}. However, the static nature of communication networks having loosely coordinated distributed control planes with no global visibility of network states and resources made it difficult to harmonize diverse RATs into a single domain. By bringing programmability in communication networks and logically centralizing the network control planes by separating it from the data forwarding plane, SDN made it possible to use a mix and match of different network equipment from different vendors. Therefore, Paper [III] proposes a HetNet architecture, as shown in Fig.~\ref{dr5}, leveraging the concepts of SDN.  

\begin{figure}[h]
\centering
\includegraphics[scale =0.32]{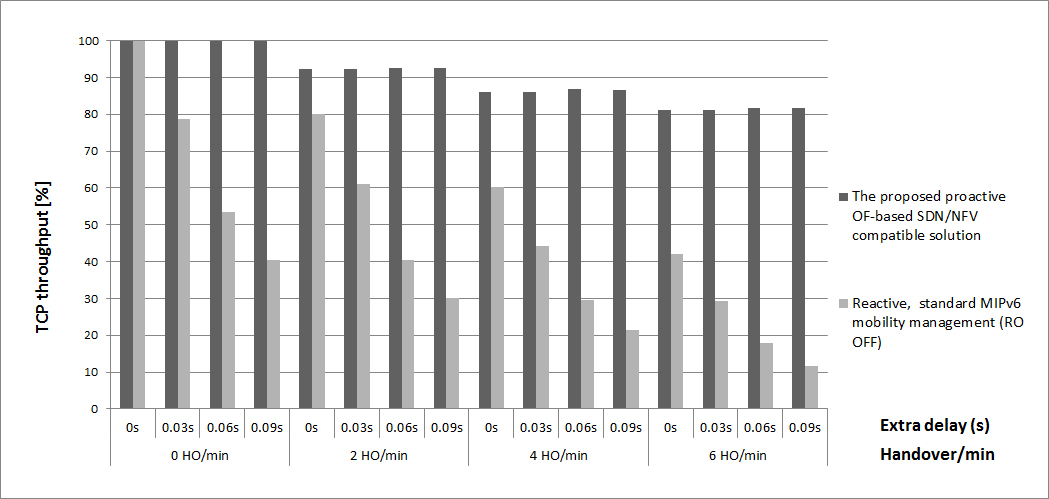}
\caption{TCP throughput of standard MIPv6-like vs. the proposed proactive SDN-based mobility management solution.}
\label{dr51}
\end{figure}

The HetNet architecture integrates different access technologies such as cognitive base stations, WLAN, and a wired network into the SDN-based centralized control platform. The paper sheds light on how resources can be dynamically allocated in such architectures. Furthermore, the paper presents a proactive SDN-based mobility management scheme between different access networks. Media Independent Handover (MIH)~\cite{4752687} scheme is used to optimize handovers among the networks. The performance evaluation and comparative analysis show that the proposed mechanisms yield better throughput and packet loss results than the standard MIP based scheme, as depicted in Fig.~\ref{dr51}. From the point of security, the proposed dynamic security tunnel management scheme drastically minimizes the signaling costs for secure IPSec tunnel establishment between backhaul devices (OpenFlow switches in our case).

\section{Security of future networks using SDN}

Paper [IV] presents a thorough study of security challenges in SDN, the proposed solutions for those challenges, and the use of SDN in strengthening network-wide security. The study also finds the potential security challenges that still expose the network to security vulnerabilities and provides future directions for strengthening security of SDNs. One of the main open challenges in SDN is that of the control channel between the control and data planes. Therefore, Paper [V] presents a security scheme for the control channel. Paper [VI] presents the security analysis of 5G networks. The security challenges that will be faced by the most important technologies used in 5G and the solutions for each of the technology are presented. The articles are summarized below.

\subsection{Security analysis of SDN}

Paper [IV] presents a study of the security challenges in SDN, the solutions proposed for those challenges, outlines the benefits of SDN for strengthening network security, and provides future directions for improving security of SDNs. Programmable networks have been prone to security challenges, and SDN is no exception. Therefore, it is highly important to investigate the loopholes and design or propose solutions to mitigate the security risks before the deployment of SDNs. For instance in active networking~\cite{916839},~\cite{1243522}, user injected programs could change the network behavior without proper authorization. If such behavior is allowed in SDNs, the security challenges will be far more detrimental for many reasons. For example, SDN has been far more accepted by the industry and academia leading to its practical deployment to serve digital services that have penetrated to our social fabric. Thus, the security of SDN circumvented by compromises in the centralized control of nearly all the network elements used by digital services can be catastrophic. Remember that SDN applications can change the network behavior through generating flow configurations and deploying them in the traffic forwarding elements. Therefore, authentication and authorization of SDN applications would constitute the basic security requirements.

The paper establishes the need of investigating security in SDNs by outlining the security challenges that existed in the previous proposals of programmable networking. Security has been the delimiting factor of previous proposals and thus new or alternative architectures have been proposed that can enable programmability of network elements securely. The predecessor of SDN, Ethane~\cite{casado2007ethane}, considered as the driving force behind the OpenFlow variant of SDN, had a specific focus on security and considered security as a subset of network management. Similarly, network-wide coherent policy deployment is another major consideration in Ethane that is propagated to SDNs. Network-wide coherent policies established through logically centralizing, the previously independent and distributed, control planes can avoid security policy contradictions in large networks.  

The centralization of the control plane and the programmability introduced by SDN are deemed highly beneficial for future networks. However, the same reasons open SDNs for security vulnerabilities. Therefore, the security challenges that arise due to these two features are properly discussed. Since SDN simplifies the network forwarding elements by shifting the control or decision-making capabilities to the controller, these data plane elements are easy to fingerprint as dependent on the controller. Thus, these elements can be easily targeted for saturation attacks besides manipulating the optional choice of using TLS for the control channel security in the OpenFlow protocol. The major security threats arising due to such vulnerabilities are discussed in the paper.

Networks that are programmable by software yield many benefits and the most promising one is quick service alteration and insertion~\cite{6702557}. In SDN, the programmable control plane can be upgraded for security by either writing a software security logic module into the control plane or integrating a security monitoring and response application through the north-bound API to the control plane. SDN facilitates network-wide coherent policies, and thus, network security can be as good as the security policies. Security policies are converged to network-wide configurations through security applications utilizing the SDN controllers. These reasons led to the concepts of \textit{\enquote{software-defined security}} and the definition of SDN as \textit{\enquote{Security-Defined Networking}}~\cite{6702553}. 

\begin{figure}[ht]
\centering
\includegraphics[scale =0.70]{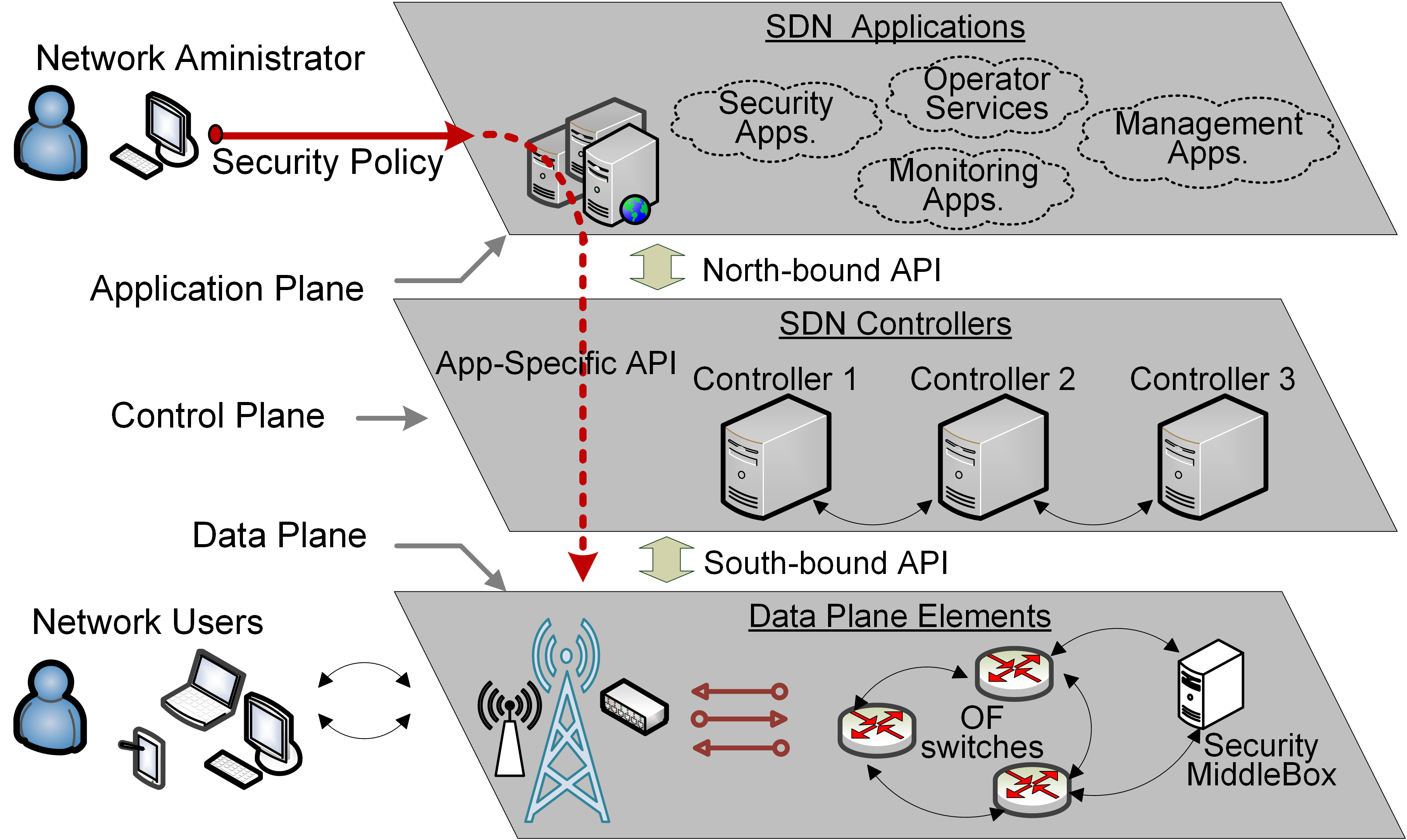}
\caption{Security policy enforcement in SDN.}
\label{dr6}
\end{figure}

The global view of the network status facilitates SDN to quickly identify a threat through a cycle of harvesting intelligence from the data plane and promptly respond to the threat. Thus, a network administrator or a security application can change the network configuration from the logically centralized control plane at run-time using the programmable APIs in the data forwarding elements to either block malicious traffic or route suspicious traffic to security middle boxes, as shown in Fig.~\ref{dr6}. Using these features of SDN, various security mechanisms are proposed to secure SDNs and develop security platforms for various types of networks. Solutions for the security of planes of SDN, and using SDN to strengthen network-wide security for different types of networks are discussed in Paper [IV], and outlined in Table~\ref{SDNSecSol}. The security solutions are tabulated in terms of the addressed threat, and the interface or SDN plane they secure.

\begin{table*}[h!]
\centering 
\caption{SDN Security Platforms} 
\begin{tabular} {|p{2.4 cm} |p{4.7 cm} |p{0.5 cm} |p{0.5  cm} | p{0.5  cm} |p{0.5  cm} | p{0.5  cm} |}
\hline\hline 
\multirow{2}{8 cm}{\textbf{Platform}} & \multirow{2}{8 cm}{\textbf{Solution}} &\multicolumn{3}{|c|}{\textbf{Target Plane}} &\multicolumn{2}{|c|}{\textbf{Interface}} \ \\ [0.5ex] 
\cline{3-7}
       &   & App.\ & Ctrl.\ & Data     \ &App-Ctrl\ &Ctrl-Data\ \\ [0.5ex] 
\hline 
FRESCO~\cite{shin2013fresco}  					& Anomaly detection and mitigation framework & & \checkmark& & \checkmark& \\ \hline
PermOF~\cite{wen2013towards}						& Permission control system for OF Apps.& & \checkmark & \checkmark &  &  \\ \hline
Assertion~\cite{beckett2014assertion}			& App debugging, Flow rules inspection &\checkmark &  & \checkmark &   &  \\ \hline
VeriFlow~\cite{Khurshid:2012:VVN:2377677.2377766}			& Verify and debug flow rules &  &  & \checkmark&  &  \\ \hline
Flover~\cite{6654813}									& Flow policy verification, identify bugs in OF programs &\checkmark & \checkmark& \checkmark& & \\ \hline
OFTesting~\cite{canini2011automating}			& App testing and debugging & \checkmark &  & & & \\ \hline
SE-Floodlight \cite{SEFloodLight}	&Role-based conflict resolution, authorization, security audit system & &\checkmark&\checkmark&\checkmark & \\ \hline
DDoSDetection \cite{braga2010lightweight}, ~\cite{6211961} & SOM-based DDoS attack detection & & \checkmark & \checkmark & &  \\ \hline
Reliable Ctrl. Placement~\cite{6134496}	& Controller reliability, switch-controller connectivity & & \checkmark & \checkmark & &\checkmark \\ \hline
Monitoring~\cite{6364688} & Data plane connectivity monitoring & & & \checkmark & & \checkmark \\ \hline
Flow rules security \cite{Porras:2012:SEK:2342441.2342466},~\cite{al2010flowchecker}	& Configuration analysis and verification, authorize applications &  & \checkmark& \checkmark&  &\\ \hline
DISCO~\cite{6838330},~\cite{6838273} & Controller availability, network monitoring &  & \checkmark & & & \\ \hline
Ctrl.-Placement \cite{heller2012controller},~\cite{6727805}		&	Controller scalability and availability & & \checkmark &  & &\checkmark \\ \hline	
Ctrl.-Reliability \cite{6573050},~\cite{tootoonchian2010hyperflow},~\cite{6662939} & Controller reliability, resilience and availability & &\checkmark & &  & \checkmark \\ \hline
CPRecovery \cite{6212011}	& Controller resilience, switch connectivity, DDoS attack & & \checkmark & \checkmark & &  \\ \hline 


\end{tabular}
\label{SDNSecSol} 
\end{table*}


The phenomena of global network state visibility and centralized control combined with programmable network elements have been identified as the potential problem solver for security challenges in many types of networks. For example, Automated Malware Quarantine (AMQ)~\cite{ONF2013DataCenterSec} uses the concepts of SDN to detect potential threats in a data center network and isolate insecure network elements to stop the spread of security threats. The security modules implemented as SDN applications monitor the network and respond through the controller to stop malicious traffic from spreading across the network. Furthermore, CloudWatcher~\cite{6459946} uses OpenFlow to monitor and inspect traffic flows to ensure security of large and dynamic clouds. Similarly, the concepts of SDN have been used to secure virtual network components. The Network Intrusion detection and Countermeasure sElection (NICE)~\cite{6419708} framework use OpenFlow to monitor and control distributed programmable virtual switches. Paper [IV] studies the use of SDN for developing such security platforms for different types of networks.

The International Telecommunication Union's Telecommunication sector (ITU-T)~\cite{itut2003secdim} has published network security recommendations to protect networks against all major security threats by defining security dimensions. These dimensions include access control, authentication, non-repudiation, data confidentiality, communication security, data integrity, availability and privacy. The study in Paper [IV] presents the security solutions for each of the seven security dimensions for SDNs, as outlined in Table~\ref{ITUT}. The study also investigates the lack of stable security solutions for SDNs with respect to each security dimension. For example, avoiding non-repudiation from SDN applications and the complexity of TLS configuration are still an open research challenges. 

Furthermore, Paper [IV] provides detailed future directions for developing SDN based security systems and security systems for future SDNs. Since the concepts of SDN such as split control-data planes architectures and network abstractions are yet to be used in communication networks, there are still many gray areas that need further investigation. For example, analysis of a system's scalability to enhance the system's security, mainly in centralized control platforms, requires further research. Similarly, many of the security problems are due to the weak notions of identity in the current Internet architecture~\cite{Rexford:2010:FIA:1810891.1810906}. The same challenge will also remain in SDN due to the lack of proper mechanisms to bind user identities to traffic flows~\cite{7226783}. Moreover, SDN can also be used to automate network security, however, security automation using SDN is yet to be explored.

\begin{table*}[ht]
\centering 
\caption{SDN Security Solutions According to ITU-T Security Recommendations} 
\begin{tabular}{|p{2.2 cm} |p{1.80 cm} |p{6.5 cm} |} 
\hline\hline 
\textbf{Security Type} &\textbf{Reference}&\textbf{Mechanism used}\ \\ [0.5ex] 
\thickhline 

{\textbf{Access}} &~\cite{wen2013towards} &Impose access control on OF apps \\ \cline{2-3}
{\textbf{Control}}&~\cite{shin2013fresco}& Enables develop security architectures for ACL \\ \cline{2-3}
										&~\cite{hinrichs2008expressing},~\cite{nayak2009resonance}& Access control policy enforcement framework \\ \thickhline

{\textbf{Authentication}} &~\cite{Porras:2012:SEK:2342441.2342466} & Role-based authentication \& authorization \\  \cline{2-3}
																	&~\cite{hinrichs2008expressing}& Authentication policies \& admission control \\ \thickhline

{\textbf{Non-}} &~\cite{yuhunag2010novel}& Uses permanent user identities (LISP) \\ \cline{2-3}
{\textbf{Repudiation}} &~\cite{6702540}	 & Uses HIP for permanent identities \\ \cline{2-3}
									&~\cite{yao2011source}			&  Source address validation of packets \\ \thickhline

{\textbf{Data}} &~\cite{jafarian2012openflow}& Random host mutation\\ \cline{2-3}
{\textbf{Confidentiality}}	&~\cite{Porras:2012:SEK:2342441.2342466}& Flow rules-legitimacy \\ \cline{2-3}
									&~\cite{6733664}	& Identity-based cryptography \\ \thickhline

{\textbf{Communication}} &~\cite{dierks2008transport} & TLS for controller-switch communication \\ \thickhline										
										
{\textbf{Data Integrity}} &~\cite{schlesingersplendid} & Traffic isolation-based integrity \\ \cline{2-3}
									&~\cite{6702540}&IPSec encapsulated security payload (ESP) \\ \cline{2-3}
									&~\cite{Khurshid:2012:VVN:2377677.2377766},~\cite{Porras:2012:SEK:2342441.2342466} & Data integrity through flow rule legitimacy.	\\ \thickhline

{\textbf{Availability}} &~\cite{6838330},~\cite{6838273} & Distribute SDN control plane\\ \cline{2-3}
	
									&~\cite{voellmy2012scalable},~\cite{cai2010system}	 & Extended processing capabilities \\ \thickhline																	
{\textbf{Privacy}} &~\cite{jafarian2012openflow} & OpenFlow random host mutation \\ \cline{2-3}
	
									&~\cite{schlesingersplendid} &  Traffic-isolation-based privacy \\ \cline{2-3}
									&~\cite{naous2009delegating}	& User-selected security procedures \\ \thickhline

\end{tabular}
\label{ITUT} 
\end{table*}

\subsection{Control channel security in OpenFlow}

Paper [V] proposes Host Identity Protocol (HIP)~\cite{5451761} based security and mobility management scheme for OpenFlow. OpenFlow uses Transport Layer Security (TLS), and the versions prior to OpenFlow version 1.0 used Secure Socket Layer (SSL)~\cite{freier2011secure}, for the control channel between the controller and data path elements~\cite{specification2015version}. However, TLS is susceptible to TCP-level attacks and changing IP addresses may tear down running sessions~\cite{7226783}. Furthermore, the use of TLS in OpenFlow is optional, hence the controller-switch communication happens in plain TCP text, leaving it open for security vulnerabilities such as reset and sequence prediction attacks. Furthermore, there are no compelling mechanisms that demonstrate the mobility of OpenFlow switches.

\begin{figure}[h!]
\centering
\includegraphics[scale =0.350]{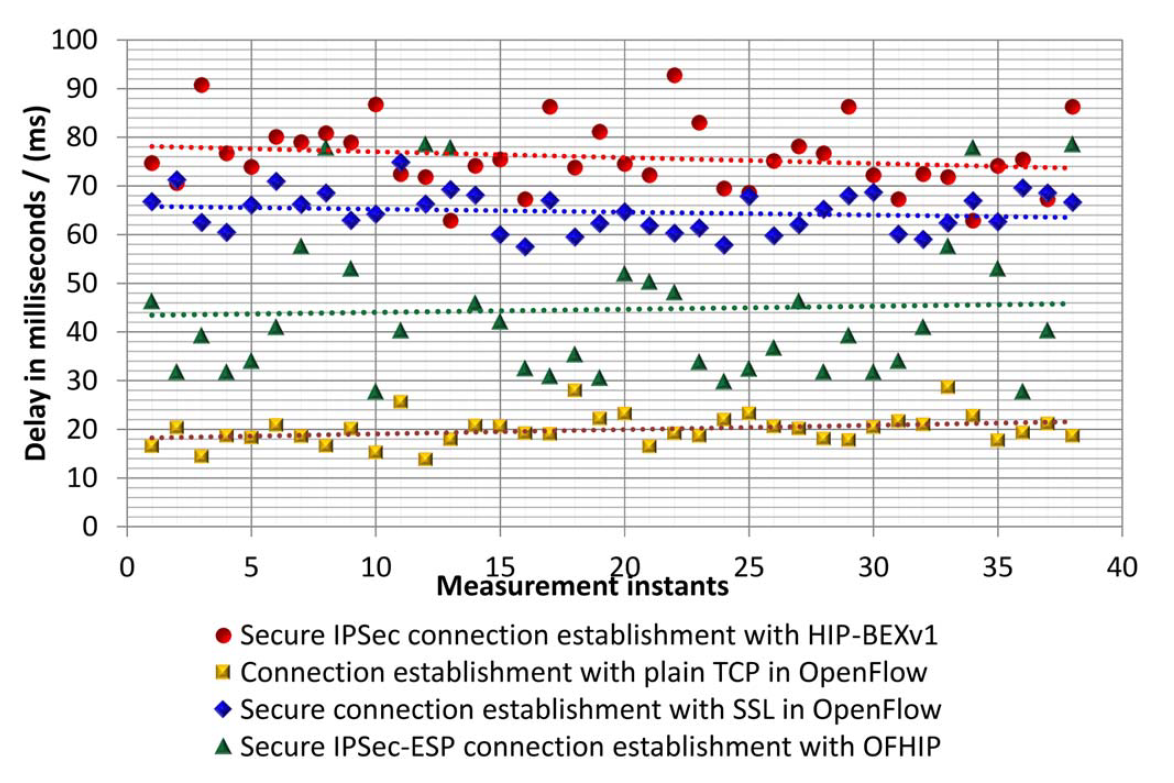}
\caption{Comparison of security schemes for control channels in SDN.}
\label{dr8}
\end{figure}

Therefore, the work presented in Paper [V] proposes a novel approach to handle security and switch mobility in OpenFlow using the global cryptographic identities introduced by HIP to replace TLS based mutual authentication. The performance results presented in Fig.~\ref{dr8} show the connection establishment delay between the two approaches. The plain TCP communication has the lowest delay but is not secure. The default (SSL/TLS) proposed in OpenFlow version 1.1.0 has an average delay of 66 ms, whereas the proposed HIP-based mechanisms has an average delay of 44 ms. The def HIP base exchange (HIP- BEX) provides the highest level of security, but introduces more delay. Further detailed results of the proposed scheme are presented in Paper [V]. 

\subsection{Security analysis of 5G networks}

Paper [VI] presents the security threat landscape of 5G networks, as depicted in Fig.~\ref{dr7}, the potential solutions for those challenges, and highlights the existing vulnerabilities to grasp attention for research towards those weaknesses. The most prominent security challenges highlighted by NGMN alliance~\cite{alliance20155g}, 5G Infrastructure Public Private Partnership (5G PPP)~\cite{5gpppsec}, and most widely discussed in the literature are:

\begin{figure}[h]
\centering
\includegraphics[scale =0.80]{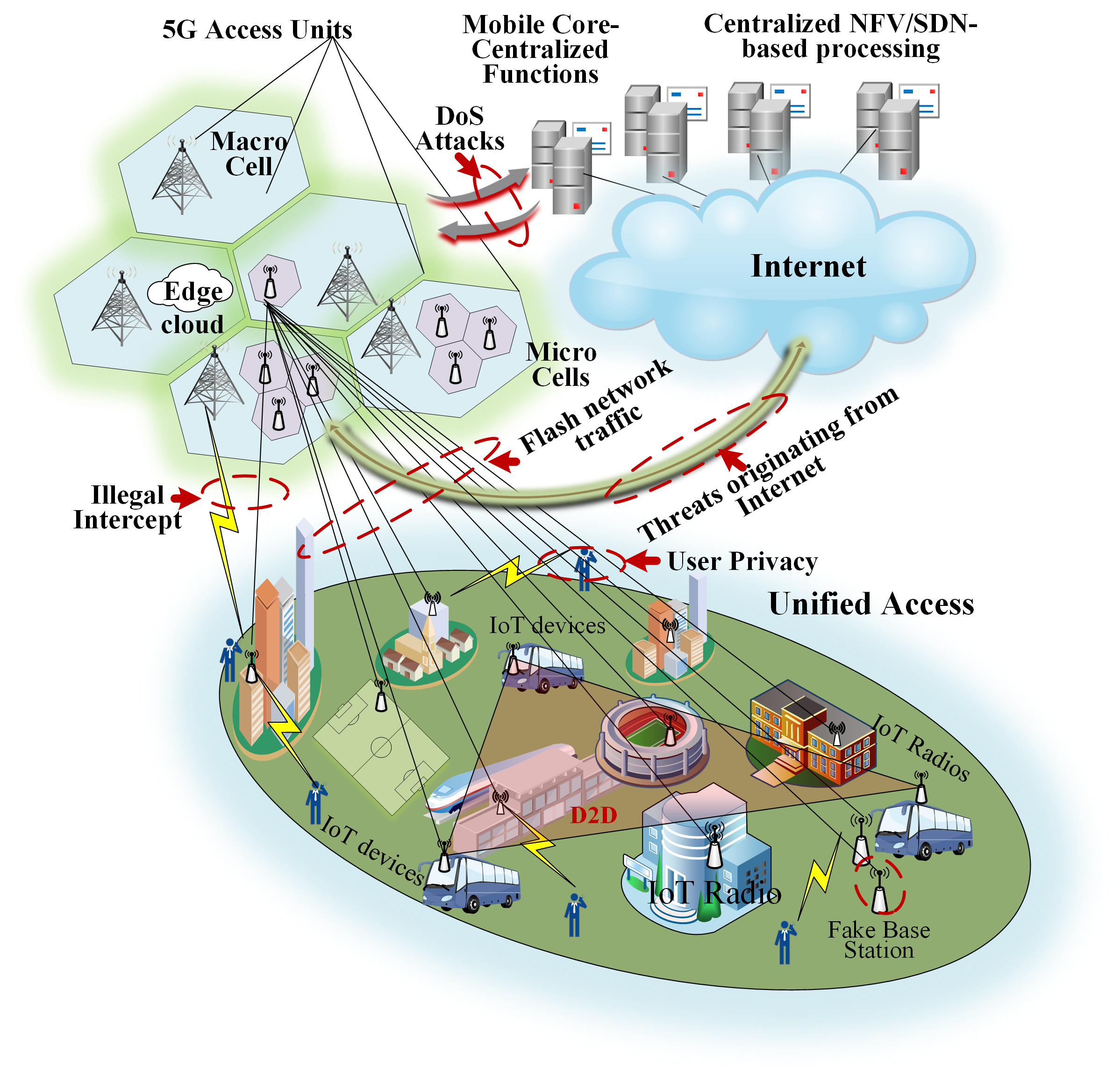}
\caption{The security threat landscape of 5G networks.}
\label{dr7}
\end{figure}

\begin{itemize}
	\item \textbf{Flash network traffic:} Generated by a huge number of end-user devices and new things (IoT) that could lead to unavailability of resources. Such traffic can also be generated by malicious users to form a Denial of Service (DoS) attack. 
\end{itemize}

\begin{itemize}
	\item \textbf{Radio interface security:} Radio interface encryption keys sent over insecure channels may lead to many security challenges, such as eavesdropping, resource stealth, etc. 
\end{itemize}

\begin{itemize}
	\item \textbf{User plane integrity:} Lacking cryptographic integrity protection for the user data plane.
\end{itemize}

\begin{itemize}
	\item \textbf{Mandated security in the network:} Service-driven constraints on the security architecture leading to the optional use of security measures. 
\end{itemize}

\begin{itemize}
	\item \textbf{Roaming security: } User-security parameters are not adjusted during roaming from between different operator networks, leading to security compromises.
\end{itemize}

\begin{itemize}
	\item \textbf{Denial of Service (DoS) attacks on the infrastructure:} Lack of mechanisms to hide visible network control elements, and unencrypted control channels.
\end{itemize}

\begin{itemize}
	\item \textbf{Signaling storms:} Distributed control systems requiring coordination, e.g. Non-Access Stratum (NAS) layer of Third Generation Partnership Project (3GPP) protocols.
\end{itemize}

\begin{itemize}
	\item \textbf{DoS attacks on end-user devices:} No proper security measures for applications, operating systems and configuration data in user devices. 
\end{itemize}

Furthermore, 5G will use a diverse set of technologies to meet the growing demands of connected devices and user data traffic. The most prominent technologies such as mobile clouds, SDN, and NFV have their own kind of security challenges that need prompt attention due to their importance in 5G. For example, Mobile Cloud Computing (MCC) inherits the security threats of cloud computing such as data and personal information theft, and privacy issues~\cite{FERNANDO201384}. Similarly, centralized control in SDN and NFV such as SDN controller and NFV hypervisor can be targeted for DoS attacks~\cite{7226783}. Therefore, it is highly important to analyze the security aspects of those technologies by finding the potential security challenges, evaluating the proposed security solutions for those challenges, and highlighting their weaknesses to grasp attention for research towards those weaknesses. Table~\ref{ttb5} presents the most important reference solutions for various attacks on different technologies used in 5G.

\begin{table}[h]
\centering 
\caption{Security solutions for various attacks in 5G technologies} 
\begin{tabular} {|p{3.7 cm} |p{2.70 cm} |p{0.60 cm} | p{0.60 cm} | p{1.1 cm} | p{0.70 cm} | p{1.15 cm} |}   
\hline\hline 
\multirow{2}{1 cm}{\textbf{Solution}} & \multirow{2}{1 cm}{\textbf{Reference}} &\multicolumn{4}{|c|}{\textbf{Target Technology}} &\multirow{2}{1 cm}{\textbf{Privacy}} \ \\ 
\cline{3-6}
                  &            & SDN & NFV & Channels & Cloud    \ & \ \\ [0.5ex] 
\thickhline

DoS, DDoS detection&\cite{braga2010lightweight},~\cite{6211961} & \checkmark &\checkmark & & & \\ \hline
Configuration &\cite{al2010flowchecker},~\cite{Khurshid:2012:VVN:2377677.2377766} & \checkmark & & & & \\ \hline
Access control&\cite{shin2013fresco},~\cite{nayak2009resonance}, ~\cite{6702551}& \checkmark &\checkmark & & \checkmark & \\ \hline

Traffic isolation&\cite{schlesingersplendid} & & \checkmark & & & \\ \hline
Link security&\cite{dierks2008transport},~\cite{7182680}& \checkmark & & \checkmark  & & \\ \hline
Identity verification&\cite{6733664},~\cite{yuhunag2010novel}, ~\cite{6702540}& & & & & \checkmark \\ \hline
Identity security&\cite{jafarian2012openflow},~\cite{Gember:2012:ELS:2396556.2396598} & & & & & \checkmark \\ \hline
Location security&\cite{7593330},~\cite{7346836}& & & & &  \checkmark  \\ \hline 
IMSI security&\cite{Norrman:2016:PIU:3021385.3021415} & & & & & \checkmark \\ \hline 
Mobile terminal security&\cite{6170530} & & & & & \checkmark \\ \hline 
Integrity verification&\cite{Khan20131278}& & & & \checkmark & \\ \hline
HX-DoS mitigation &\cite{6354861}& & & & \checkmark & \\ \hline
Service access Control&\cite{7064902}& & & & \checkmark & \\ \thickhline 
\hline

\end{tabular}
\label{ttb5}
\end{table}									

5G will connect every aspect of the society through communication networks. Hence, user privacy will be another pressing challenge in 5G networks~\cite{kumar2018user},~\cite{7980650}. Henceforth, the paper provides an overview of the existing and novel security methodologies to cope with the security challenges in each of these technologies, as well as possible solutions for maintaining user privacy. The work presented in Paper [VI] also sheds light on the standardization efforts on security in 5G. The security standardization for 5G is still under the drafting phase and many key organizations are contributing for the rapid development of security technologies and their standards. The standardization activities are also highlighted in Table~\ref{table_stand}.


\begin{table*} [h]
\renewcommand{\arraystretch}{1.0} 
\caption{Security activities of various standardization bodies }
\label{table_stand} 
\centering 
\begin{tabular}{|p{2.3 cm} |p{2.1 cm} |p{3 cm} |p{3.2 cm} |} 
\hline 
\bfseries Standardization bodies &  \bfseries Workgroups &  \bfseries Major security areas in focus &  \bfseries Milestones  \\  
\hline
3GPP & Service and System Aspects Security Group (SA3)  &Security architecture, RAN security, authentication mechanism, subscriber privacy, network slicing &  TR 33.899: study on the security aspects of next generation systems, TS 33.501: security architecture and procedures for 5G System  \\ \hline
5GPPP & 5GPPP Security WG  &Security architecture, subscriber privacy, authentication mechanism &  5G PPP Security Landscape-(White Paper) June 2017 \\ \hline
IETF & I2NSF, DICE WG, ACE WG, DetNet WG & Security solutions for massive IoT devices in 5G, user privacy, Network Security Functions (NSFs)  & RFC 8192, \ RFC 7744, Deterministic Networking (DetNet) Security Considerations   \\ \hline
NGMN & NGMN P1 WS1 5G Security Group & Subscriber privacy, network slicing, MEC security  & 5G security recommendations: Package 1 and 2,\ and 5G security: Package 3
 \\ \hline
ETSI & ETSI TC CYBER, ETSI NFV SEC WG  & Security architecture, NFV security, MEC security, privacy  & ETSI GS NFV-SEC 010,\ ETSI GS NFV-SEC 013\, ETSI GS NFV-SEC 006 and ETSI GS MEC 009  \\ \hline
\hline

\end{tabular}
\end{table*}

\chapter{Conclusion and future work}
This chapter summarizes the thesis in conclusion, highlighting the contributions and main results. Furthermore, existing problems in the scope of the thesis are described and future research directions are presented.

\section{Conclusion}


SDN lays down the foundation for flexible and adaptable communication network architectures by separating the network control plane from the data forwarding plane. The control plane is logically centralized and implemented in software that oversees and controls the simplified data plane through programmable interfaces. The idea of SDN is architected by OpenFlow in a three tier architecture comprising the application plane, the control plane, and the data plane. The control plane is capable of changing the data forwarding behavior in the data plane according to requirements of applications in the application plane. For example, OpenFlow applications can request network states such as flow table values or packet counter values from OpenFlow switches through the OpenFlow controller. OpenFlow applications can make decisions based on some algorithms, e.g. for load balancing or traffic inspection, and deploy those decisions in the data plane through the controller. SDN, thus, enables new services to manipulate the network and utilize network resources as required by users. 

This thesis evaluated the potential of SDN in future networks from two perspectives: first, investigating the potential of SDN for intelligently sharing resources and enabling heterogeneity in future wireless networks. A centralized control plane approach for multiple access technologies is proposed and evaluated with practical experiments. Second, this thesis presented the analysis of security of SDN and future networks such as 5G that uses SDN as one of the main enabling technologies. The security challenges in SDN and the solutions for those challenges are studied and future directions for increasing the security of SDNs are outlined. The main security challenges in 5G are discussed and the potential solutions for those challenges, including enhancing security by leveraging SDN, are studied.

Cognitive networking aimed at making communication networks intelligent enough to automatically respond and fulfill user needs under the constraints of available network resources. However, the stringent nature of existing networks having hardware based network functions delimited the use of cognitive networking only to the radio part. Thus, the concepts of SDN and cognitive networking have been integrated to fulfill the promise of cognitive networking in terms of network flexibility and adaptability beyond the radio part. The experimental evaluations show that the proposed centralized control framework, using the currently implemented SDN architecture in the form of OpenFlow, has capabilities to provide the necessary dynamism. The performance improvement has been demonstrated in terms of throughput, packet loss, and signaling costs, etc.

However, SDN has its own challenges and, among those, security is on the forefront. Therefore, this thesis discussed the security weaknesses and threats in SDN, possible solutions for such vulnerabilities, and highlighted the unexplored security limitations of SDN for future research. Since 5G will connect most aspects of the human life through the communication infrastructure, security issues must be highlighted early to seek solutions through grasping research attention. This thesis evaluated the security vulnerabilities in 5G and the technologies that 5G will use such as cloud computing, SDN, and NFV, and presented solutions to those challenges. Having the SDN control channel being recognized as the most critical interface in the network, this thesis also presented a novel approach that will not only provide security to the control channel, but will also help mobility in SDNs.

\section{Discussion and future work}

SDN will play a major role in next generation communication networks due to the flexibility it offers through programmability and simplified network control and management it provides through global network state visibility. Decoupling the network control plane from the data forwarding plane and logically centralizing the control plane has been widely accepted as the way forward, and that is the true potential of SDN. Therefore, the main objectives of this thesis have been to investigate the potential of SDN in its current form for the very next generation of networks such as the fifth generation (5G) wireless networks, and how SDN can fulfill the requirements of future networks. The thesis outlines the benefits of using SDN in future networks through practical or experimental evaluations, highlights the potential challenges in terms of security, discusses the security solutions for those challenges alongside the remaining loopholes and proffers future directions for security.

However, there are still many open questions that need further investigation. For example, the concepts of cognitive networking have been proposed mainly to enable network automation and to eliminate or minimize the need for human intervention. Cognitive radios have achieved a level of automation using Software defined radios to tune-in to the available frequency channels. Full automation that detects user service needs and then decides the selection of frequency bands based on the service requirements is an interesting future research question. Furthermore, a fully automated network might take a top down approach where all the elements of a network are synchronized. Synchronized here means that the network, from application to the data forwarding plane, adjusts itself according to end-to-end goals of users, services and network operators while optimizing the available resources. 

Using the current implementation of SDN, i.e. OpenFlow, with the centralized controller for automated cognitive network will raise further questions also. For example, rapid changes in user behavior or network operating conditions will require the SDN controller to instantly adjust the network environment accordingly. In situations where changes are frequent, such as frequency hopping, the SDN controller might be consumed to a level where it will not be able to respond to network-wide goals or network configurations for other services. This might lead to scalability challenges due to the involvement of the centralized controller. Hence, relegating control functionalities or distributing the control plane functionalities for cognitive networking is another interesting area that needs further research.

In the testbed setup for experiments of SDN-based cognitive networking, the Wireless Open-Access Research Platform (WARP) platforms had limitations in terms of bandwidth. For example, on a single channel, the maximum achievable bandwidth remained well below what 5G offers or even what existing cellular networks offer. Therefore, with using multiple frequency channels, the bandwidth still remained very low, for example, approximately 6 Mbps in our experiments. This is attributed to the limitations in WARP platforms. The main focus of the work presented in the thesis, however, is the dynamic use of spectrum resources in a HetNet environment using the SDN-based control platform. Hence, the results must be seen from perspectives other than the low bandwidth, such as the benefits of centralized control of radio resources. 

As an interesting future research topic, cognitive networking needs further investigation for automation in future communication networks. IoT will play a major role in the near future and the number of IoT devices is expected to grow in billions. Automatic resource provisioning for IoT, thus, will be the key requirement of future networks. How to enable network nodes and network segments to cooperate in order to grow and shrink in capacity at run-time will be a key research area. Using SDN to dynamically place network functions and automatically configure network equipment; and using cognition to sense the needs and behavior of IoT devices with different capabilities demands further research.  

SDN has many benefits such as innovation in communication networks, simplified network management and reduced costs. However, there are still many open research challenges that need further investigation. For example, the degree of involvement of SDN controller in network configurations and flow setups; and the physical and logical or virtual placement of controllers in large networks are some of the open questions regarding the SDN controller. The idea of decoupled architecture also has challenges, however. For example, fingerprinting the decoupled architecture, and thus the control and data planes in the network, is comparatively easy in which either of them can be targeted for security attacks. However, SDN is inherently less vulnerable to security threats than the previous proposals for programmable network architectures, such as active networking. Therefore, the use of the concepts of SDN will continue, not necessarily in its current implementation in the form of OpenFlow, due to the many benefits it offers.

%
%

%

\addcontentsline{toc}{chapter}{\bibname}
\fontsize{9}{9}\selectfont

%
%
%

\bibliographystyle{IEEEtran}
\bibliography{IEEEbibl}



\newpage 

Original publications for the thesis:



\begin{enumerate}[label=\Roman*, widest=VIII]
\fontsize{9}{9}\selectfont


\item    Ahmad, I., Namal, S., Ylianttila, M., \& Gurtov, A. (2015, July). Towards software defined cognitive networking. In New Technologies, Mobility and Security (NTMS), 2015 7th International Conference on (pp. 1-5). IEEE. 
\label{I}

\item   Namal, S., Ahmad, I., Saud, S., Jokinen, M., \& Gurtov, A. (2016). Implementation of OpenFlow based cognitive radio network architecture: SDN\&R. Wireless Networks, 22(2), 663-677. 
\label{II}

\item    Ahmad, I. Liyanage, M., Bokor, L., Ylianttila, M., \& Gurtov, A. (2018). "Enabling Heterogeneity in 5G Using the Concepts of SDN", Manuscript. 
\label{III}

\item   Ahmad, I., Namal, S., Ylianttila, M., \& Gurtov, A. (2015). Security in software defined networks: A survey. IEEE Communications Surveys \& Tutorials, 17(4), 2317-2346.
\label{IV}

\item   Namal, S., Ahmad, I., Gurtov, A., \& Ylianttila, M. (2013, November). Enabling secure mobility with openflow. In Future Networks and Services (SDN4FNS), 2013 IEEE SDN for (pp. 1-5). IEEE.
\label{V}

\item  Ahmad, I., Kumar, T., Liyanage, M., Okwuibe, J., Ylianttila, M., \& Gurtov, A. (2018, March). Overview of 5G Security Challenges and Solutions. In IEEE Communications Standards Magazine, vol. 2, no. 1, pp. 36-43, MARCH 2018. IEEE.
\label{VI}


\end{enumerate}


Original publications are not included in the electronic version of the dissertation.\\

Published thesis available on University of Oulu publication channel, Jultika, on the link: http://jultika.oulu.fi/files/isbn9789526219516.pdf

\end{document}